%% file: main.tex
\documentclass[12pt]{article}
\usepackage[margin=1in]{geometry}

\usepackage{amsmath}
\usepackage{amssymb}
\usepackage{graphicx}
\usepackage{booktabs}
\usepackage{multirow}

\usepackage{hyperref}

\usepackage{xcolor}

\usepackage{natbib}

\usepackage{subcaption}

\usepackage{algorithm}
\usepackage{algpseudocode}

\newcommand{\myurl}[1]{\href{#1}{\texttt{\nolinkurl{#1}}}}

\DeclareMathOperator*{\argmin}{arg\,min}

\DeclareMathOperator*{\KL}{KL}
\DeclareMathOperator{\E}{E}

\DeclareMathOperator*{\upc}{\overset{+}{\approx}}
\DeclareMathOperator{\diag}{diag}

\newcommand{\R}{\textsf{R}\ }
\newcommand{\pck}[1]{\textsf{#1}}

\newcommand{\fn}[1]{\texttt{#1}}
\newcommand{\trp}{^{\prime}}

\title{Variational Inference for Variable Selection in Scalar-on-Function Regression}
\author{Ana Carolina da Cruz$^{1*}$ \and \ Camila P.\ E.\ de Souza$^1$ \and \ Pedro H.\ T.\ O.\ Sousa$^2$ \\  
\normalsize $^1$Department of Statistical and Actuarial Sciences, University of Western Ontario, Canada\\
\normalsize $^2$Department of Statistics, Federal University of Paraná, Brazil \\
\normalsize$^*$Corresponding author: adacruz@uwo.ca}
\date{}

\begin{document}

\maketitle

\begin{abstract}
In practical regression applications, multiple covariates are often measured, but not all may be associated with the response variable. Identifying and including only the relevant covariates in the model is crucial for improving prediction accuracy. In this work, we develop a variational inference approach for estimation and variable selection in scalar-on-function regression, involving only functional covariates, and in partially functional regression models that also include scalar covariates. Specifically, we develop a variational expectation–maximization (VEM) algorithm, with a variational Bayes procedure implemented in the E-step to obtain approximate marginal posterior distributions for most model parameters, except for the regularization parameters, which are updated in the M-step. Our method accurately identifies relevant covariates while maintaining strong predictive performance, as demonstrated through extensive simulation studies across diverse scenarios. Compared with alternative approaches, including BGLSS (Bayesian Group Lasso with Spike-and-Slab priors), grLASSO (group Least Absolute Shrinkage and Selection Operator), grMCP (group Minimax Concave Penalty), and grSCAD (group Smoothly Clipped Absolute Deviation), our approach achieves a superior balance between goodness-of-fit and sparsity in most scenarios. We further illustrate its practical utility through real-data applications involving spectral analysis of sugar samples and weather measurements from Japan.



\end{abstract}

\section{Introduction}

Functional data analysis (FDA) has become increasingly popular in recent decades, largely due to technological advancements that have enabled the collection of complex, high-dimensional data naturally represented as functions, such as growth curves and signal processing data \citep{ramsay1991, ramsay2005, ferraty2006}. Due to its high-dimensionality and complex-data structure, modelling functional data poses significant challenges in the estimation and interpretation, requiring the development of approaches that are scalable and computationally efficient \citep{ramsay2009, wang2016}.

A particularly active and growing area within FDA is functional regression \citep{goldsmith2017,ghosal2020,yeh2023,huo2023,zhou2023,sousa2024,sun2025}. Like its non-functional counterpart, functional regression is used to describe the relationship between a single or multiple covariates and a response variable, particularly when either the response or one or more covariates are functional. Functional regression models are typically categorized into three main types: functional response regression (function-on-scalar), where the response variable is functional and the covariates are scalar variables; functional predictor regression (scalar-on-function), where the response variable is scalar and the covariates are functional variables; and function-on-function regression, where both the response variable and covariates are functional \citep{ramsay2005, morris2015}.

In practical regression applications, multiple covariates are often measured, but not all may be associated with the response variable. Identifying and including only the relevant covariates in the model is crucial for improving prediction accuracy \citep{hastie2009}. As a result, variable selection in functional regression has gained significant attention over the past decade.
In this work, we focus on variable selection in scalar-on-function regression (SoFR), where multiple functional covariates are observed, but not all may be associated with the scalar response. This type of functional regression has been widely applied across diverse domains, including the analysis of mass spectrometry data \citep{zhao2012, di2023}, physical activity \citep{jadhav2022, zoh2024}, and meteorological measurements \citep{collazos2016, mbina2025}. 
In this context, we define variable selection as the identification of functional covariates that have an effect on at least a small region of the domain; that is, we aim to remove variables that exhibit no effect on the response over their entire domain. 

Inference for SoFR models has been conducted using both frequentist and Bayesian approaches. Here, we focus on Bayesian inference methods as they offer advantages over frequentist approaches, such as the ability of direct quantification of model uncertainty through the construction of credible bands and incorporation of prior knowledge into the model via the specification of parameter prior distributions. Variational inference (VI) has gained popularity in Bayesian inference for its efficiency and scalability \citep{jordan1999, blei2017}, offering substantial computational advantages over Markov chain Monte Carlo (MCMC), particularly in functional data analysis \citep{goldsmith2016, huo2023, xian2024, dacruz2024}. For a review on frequentist techniques applied to SoFR, we refer the reader to \citet{reiss2017} and \citet{aneiros2022}.

Among Bayesian approaches proposed for SoFR, several methods focus exclusively on estimation without incorporating variable selection. Variational inference (VI) methods include the published works of \citet{goldsmith2011} and \citet{garcia2024}, as well as the preprints by \citet{mclean2013} and \citet{meyer2023}. With the exception of \citet{garcia2024}, these contributions consider only a single functional covariate. \citet{goldsmith2011} developed a variational Bayes (VB) algorithm that accounts for measurement error in a functional covariate represented via functional principal components (FPCA). In the context of mixture models, \citet{garcia2024} proposed a VB framework for unsupervised Bayesian classification with mixing probabilities depending on scalar and functional covariates. \citet{mclean2013} introduced a VB algorithm for functional generalized additive models (FGAM) accommodating a sparsely and irregularly observed functional covariate with measurement error, while \citet{meyer2023} adapted the Akaike information criterion (AIC) to the functional regression setting to select non-functional components in mixed-effects models. MCMC procedures have likewise been developed for estimation in SoFR without performing variable selection, including the recent contributions by \citet{zoh2024} and \citet{kang2024}. \citet{zoh2024} addressed complex measurement error structures via instrumental variables, while \citet{kang2024} proposed an MCMC framework for scalar-on-function regression with spatial random effects.

All variable selection approaches found in the context of SoFR rely on MCMC, for which we cite the works by  \citet{zhu2010} and \citet{banik2022}. \citet{zhu2010} proposed an approach for variable selection of functional predictors in functional generalized linear regression with binary responses. The functional covariates are represented using FPCA and are identified as relevant (or not) via latent variables included in the prior distributions of their corresponding functional coefficients. In the same functional generalized linear regression framework as \citet{zhu2010}, \citet{banik2022} adopt a spike-and-slab prior on the functional coefficients, combined with a Pólya-gamma distribution, to select relevant covariates. Although BGLSS (Bayesian group LASSO with spike-and-slab priors) \citep{ghosh2015} was not originally developed for functional regression, it is an MCMC-based method that has been adapted and applied to variable selection in functional regression settings \citep{sousa2024fosr}.

The literature review above reveals a gap in VI methods to perform variable selection in SoFR models. We fill this gap by developing a novel VI algorithm for inference and variable selection in SoFR. Our approach builds upon the idea of considering latent variables to perform variable selection as in \citet{sousa2024fosr}. However, our method is fundamentally different from theirs in the sense that we consider a scalar response variable and multiple functional covariates instead of a functional response and scalar covariates. Additionally, we develop a variational expectation-maximization (VEM) algorithm \citep{neal1998, dimitris2008} to perform inference instead of an MCMC approach via a Gibbs sampler. Our VEM approach consists of a VB algorithm in the E-step, where marginal posterior estimates are obtained for most of the model parameters, except for the regularization parameters, which are estimated in the M-step. To the best of our knowledge, our work is the first to propose a model-based VI approach for variable selection in SoFR.

The remaining of this paper is organized as follows. In Section \ref{sec:model_sofr}, we introduce our proposed method to perform estimation and variable selection in scalar-on-function regression under a Bayesian framework. We present a quick overview of VB and VEM approaches in Sections \ref{sec:vb_overview} and \ref{sec:vem} and derive our proposed VEM algorithm in Section \ref{sec:vb_sofr}. In Section \ref{sec:sim_sofr}, we describe the simulation studies design and present the performance of our method under each one of them in Section \ref{sec:sim_res_sofr}. Section \ref{sec:real_data_sofr} illustrates the application of our proposed method to the sugar spectra and the Japan weather datasets. A discussion about the implications of our proposed method is provided in Section \ref{sec:conclusion_sofr}.

\section{Methods}

In Section \ref{sec:model_sofr}, we first introduce our proposed Bayesian hierarchical model to perform estimation and variable selection in scalar-on-function regression. We expand the functional covariates and functional coefficients using basis functions and employ latent indicator random variables to allow for adaptive covariate selection. Then, in Sections \ref{sec:vb_overview} and \ref{sec:vem} we present an overview of VB and VEM algorithms followed by the derivation of the update equations given in Section \ref{sec:vb_sofr}.

\subsection{Model}
\label{sec:model_sofr}

Let $Y_i, \; i = 1, \dots, n$, be a scalar response, and let $X_{ij}(t),\; j = 1, \dotsc, p$, denote the $j$th functional covariate for the $i$th observation, evaluated at point $t$ (such as time or another quantity in a continuum) within the domain $\mathcal{T}_j$. Each covariate $X_{ij}(t)$ is assumed to be known and smooth. The scalar-on-function regression model is then given by
\begin{equation}
\label{eq_sofr_model}
Y_i = \beta_0 + \sum_{j=1}^p \int_{\mathcal{T}_j} X_{ij}(t)\beta_j(t)dt + \varepsilon_i,
\end{equation}
where $\beta_0$ is the intercept, $\beta_j(t)$ is the unknown smooth effect of $X_{ij}(t)$ on the response at evaluation point $t$, and $\varepsilon_1, \dotsc, \varepsilon_n$ are independent and identically distributed Gaussian random errors with a mean of zero, a variance of $\sigma^2$, and are independent of the $X_{ij}(\cdot)$s. For notational simplicity, we assume that all predictors are defined on the same domain $\mathcal{T}$, so that $\mathcal{T}_j = \mathcal{T},$ for all $j$. Because our primary interest is in estimating the functional coefficients, in what follows, we omit $\beta_0$ from Equation \eqref{eq_sofr_model} for ease of presentation. Note that the intercept can be computed after model estimation as, $\hat\beta_0 = \bar{Y} - \sum_{j=1}^p \int_{\mathcal{T}} \bar{X}_{j}(t)\hat\beta_j(t)dt$, where $\bar{Y}$ is the average response, $\bar{X}_{j}(t)$ is the functional average of predictor $X_{j}(t)$ at evaluation point $t$ and $\hat\beta_j(t)$ denotes its corresponding estimated functional coefficient.

The model in Equation \eqref{eq_sofr_model} can also include scalar covariates, leading to the so-called partially functional regression model. Our work focuses on a VI approach for variable selection in SoFR, for which we present, in what follows, the hierarchical model and VEM algorithm. A separate simulation study (Simulation Study 3, Section \ref{sec:sim_sofr}) considers the joint selection of scalar and functional covariates, with the corresponding partially functional VEM algorithm provided in Section 1 of the Supplementary Material. 

Our variable selection approach introduces latent Bernoulli random variables $Z_j$ to identify covariates for which $\beta_j(t) = 0,$ for all $t \in \mathcal{T}$, yielding the model
\begin{equation}
\label{eq_sofr_model_vs}
Y_i = \sum_{j=1}^p \int_{\mathcal{T}} X_{ij}(t)\underbrace{Z_j\beta_j(t)}_{\xi_j(t)}dt + \varepsilon_i,
\end{equation}
where $Z_j$ determines whether the $j$th functional covariate is relevant for explaining $Y_i$, with $P(Z_j = 1) = \theta_j$. Therefore, in what follows, we will need to differentiate $\beta_j(t)$ and $\xi_j(t) = Z_j\beta_j(t)$. We define $\beta_j(t)$ as the $j$th partial functional coefficient and  $\xi_j(t)$ as the $j$th final functional coefficient upon selection. Our variable selection approach, commonly used in Bayesian variable selection frameworks, involves employing sparsity-inducing priors, such as latent binary variables \citep{ormerod2017, zhang2019, sousa2024fosr} or spike-and-slab priors \citep{rao2005, ghosh2015, huo2023}. We adopt the former by using latent variables with Beta--Bernoulli structure to shrink non-relevant functions $\xi_j(\cdot)$ toward zero, while estimating the remaining functions with non-zero effects.

Since we assume that the functional covariates are known and smooth, we first represent them via basis function expansions, with the choice of basis determined by the characteristics of the data. In this work, we consider cubic B-splines and estimate the corresponding basis coefficients using regression splines implemented via the \fn{smooth.basis} function in the \R package \pck{fda} \citep{fda2024}. Then, we proceed inferring the parameters of the model in Equation \eqref{eq_sofr_model_vs} considering the coefficients of the basis expansion of the functional covariates fixed. Each functional coefficient is expanded using the same basis functions as its associated covariate, which facilitates the computation of inner products between basis matrices and allows for efficient implementation through the \fn{inprod} function in the \R package \pck{fda}. The basis coefficients for these expansions, together with the remaining model parameters, are then estimated using the proposed VEM algorithm.


This representation of both functional covariates and functional coefficients in terms of basis functions permits the functional regression model in Equation \eqref{eq_sofr_model_vs} to be reformulated as a multivariate linear regression model in the basis space. Let $\mathbf{B}_{j}(t) = (B_{j1}(t), \dotsc, B_{jK}(t))\trp, \; j = 1, \dotsc, p$ and $t \in \mathcal{T}$ be the vector of $K$ known cubic B-splines basis functions for each functional covariate $j$, same across observations. Then, each functional covariate and its corresponding functional coefficient can be represented as
\begin{equation}
\label{eq:X_smooth}
X_{ij}(t) = \sum_{k=1}^{K}a_{ijk}B_{jk}(t) = \mathbf{A}_{ij}\trp \mathbf{B}_{j}(t),
\end{equation}
and
\begin{equation}
\label{eq:beta_basis}
\beta_j(t) = \sum_{k=1}^K b_{kj}B_{jk}(t) = \mathbf{b}_{j}\trp\mathbf{B}_{j}(t),
\end{equation}
where $\mathbf{A}_{ij}= (a_{ij1}, \dotsc, a_{ijK})\trp$ is the vector of $K$ known basis coefficients for functional covariate $j$, observation $i$,  and $\mathbf{b}_{j} = (b_{1j}, \dotsc,  b_{Kj})\trp$ is the vector of $K$ unknown basis coefficients for the $j$th functional coefficient. Then, using Equations \eqref{eq:X_smooth} and \eqref{eq:beta_basis} the functional regression model in Equation \eqref{eq_sofr_model_vs} can be expressed in the following form:
\begin{align}
\label{eq:model_linear}
    Y_i &= \sum_{j = 1}^p\int_{\mathcal{T}} \mathbf{A}_{ij}\trp\mathbf{B}_{j}(t)\big(Z_j\mathbf{B}_{j}(t)\trp \mathbf{b}_{j}\big) dt + \varepsilon_i\nonumber\\
    &= \sum_{j=1}^p\mathbf{A}_{ij}\trp Z_j\left(\int_{\mathcal{T}}\mathbf{B}_{j}(t)\mathbf{B}_{j}(t)\trp  dt \right)\mathbf{b}_{j} + \varepsilon_i\nonumber\\
    &= \sum_{j=1}^p Z_j(\mathbf{W}_{ij}\trp\mathbf{b}_j) + \varepsilon_i = \mathbf{W}_i\trp\Gamma\mathbf{b} + \varepsilon_i,
\end{align}
or, in matrix form,
\begin{equation*}
\mathbf{Y} = W\Gamma\mathbf{b} + \boldsymbol{\varepsilon},
\end{equation*}
where $\mathbf{Y} = (Y_1, \dotsc, Y_n)\trp$, $\Gamma = \diag{\boldsymbol{\gamma}}$ is a $Kp \times Kp$ diagonal matrix, with $\boldsymbol{\gamma} = \mathbf{Z}\otimes\boldsymbol{1}_K$, $\mathbf{Z} = (Z_1, \dotsc, Z_p)\trp$, $\mathbf{b} = (\mathbf{b}_1\trp, \dotsc, \mathbf{b}_p\trp)\trp$ is a $Kp$ vector of unknown basis coefficients, $W = (\mathbf{W}_1\trp, \dotsc, \mathbf{W}_n\trp)\trp$ is a $n \times Kp$ matrix, with $\mathbf{W}_i = (\mathbf{W}_{i1}, \dotsc,  \mathbf{W}_{ip})\trp$, $\mathbf{W}_{ij} =  J_{j}\trp\mathbf{A}_{ij}$, $J_{j} = \int_{\mathcal{T}}\mathbf{B}_{j}(t)\mathbf{B}_{j}(t)\trp  dt$ is a $K \times K$ cross product matrix, $\boldsymbol{\varepsilon} = (\varepsilon_1, \dotsc, \varepsilon_n)$ is the vector of error terms and $\otimes$ represents the Kronecker product operator. The cross product matrix $J_j$ can be calculated by using numerical integration or as described in \citet{kayano2009}.

Based on Equation \eqref{eq:model_linear} and additional assumptions described above, we define the proposed Bayesian hierarchical model as
\begin{align}
\label{eq:hier_sofr}
Y_i \mid \mathbf{Z}, \boldsymbol{\beta}, \sigma^2 &\sim \mathcal{N}\left(\mathbf{W}_i\trp\Gamma\mathbf{b}, \sigma^2\right); \nonumber \\
b_{kj} \mid \sigma^2, \tau_{kj}^2 &\sim \mathcal{N}(0, \tau_{kj}^2\sigma^2); \nonumber \\
Z_{j} \mid \boldsymbol{\theta} &\sim \mathrm{Bernoulli}(\theta_{j}); \nonumber \\
\theta_{j} &\sim \mathrm{Beta}(0.5,0.5);\\
\tau_{kj}^2 &\sim \mathrm{Exponential}\left(\frac{\lambda_j^2}{2}\right); \nonumber \\
\sigma^2 &\sim \mbox{Inverse-Gamma}(\delta_1, \delta_2), \nonumber 
\end{align}
where $\lambda^2_j$ and $\tau_{kj}^2$ control the regularization of the $k$th basis coefficients associate to the $j$th functional coefficient. These choices of priors have been previously considered in the literature by \citet{casella2010} and \citet{sousa2024fosr}. Different strategies can be used to determine the degree of regularization of the $b_{kj}$s, including placing prior distributions (e.g., Gamma priors) on the $\lambda^2_j$s or estimating them directly via point estimation within a VEM framework. We conducted a preliminary evaluation of both approaches using simulated data (results not shown). This assessment indicated that the prior-based approach was sensitive to the choice of hyperparameters and made it difficult to achieve the desired level of regularization in practice. As a result, we decided to adopt a point-estimation strategy where the regularization parameters are estimated in the M-step of a VEM algorithm. In the following sections, we provide a brief overview of VB and VEM procedures and then describe the steps of our proposed algorithm.

\subsection{Overview of variational Bayes}
\label{sec:vb_overview}

Bayesian inference often relies on MCMC, which generates samples from the posterior but can be computationally demanding. VI provides an alternative by reformulating posterior approximation as an optimization problem rather than a sampling task \citep{jordan1999, wainwright2008, blei2017}. Let $\boldsymbol{\theta}$ denote the model parameters, $\boldsymbol{y}$ the observed data, and $q$ a density function in the variational family $\mathcal{Q}$. The optimal approximation is then defined as
\begin{equation}
q^{*}(\boldsymbol{\theta}) = \argmin_{q \in \mathcal{Q}} \KL\!\left(q(\boldsymbol{\theta}) \,\|\, p(\boldsymbol{\theta} \mid \boldsymbol{y})\right).
\end{equation}

The KL divergence can be written as
\begin{align}
\KL\!\left(q(\boldsymbol{\theta}) \,\|\, p(\boldsymbol{\theta} \mid \boldsymbol{y})\right)
&= \E_q \!\left[\log \frac{q(\boldsymbol{\theta})}{p(\boldsymbol{\theta} \mid \boldsymbol{y})}\right]  \\
&= \E_q[\log q(\boldsymbol{\theta})] - \E_q[\log p(\boldsymbol{\theta} \mid \boldsymbol{y})] \nonumber \\
&= - \Big(\E_q[\log p(\boldsymbol{\theta}, \boldsymbol{y})] - \E_q[\log q(\boldsymbol{\theta})]\Big) + \log p(\boldsymbol{y}),\nonumber
\end{align}
where expectations are taken with respect to $q(\boldsymbol{\theta})$. Since $\log p(\boldsymbol{y})$ does not depend on $q$, minimizing the KL divergence is equivalent to maximizing the evidence lower bound (ELBO) \citep{blei2017}, given as
\begin{equation}
\mathrm{ELBO}(q) = \E_q[\log p(\boldsymbol{\theta}, \boldsymbol{y})] - \E_q[\log q(\boldsymbol{\theta})].
\end{equation}
A common choice of variational family $\mathcal{Q}$ is the mean-field variational family, which assumes independence among parameters, with the variational distribution being factorized as
\begin{equation*}
    q(\boldsymbol{\theta}) = \prod_j q_j(\theta_j).
\end{equation*}

This factorization makes computations more tractable. Optimization under this family is typically performed using the coordinate ascent variational inference (CAVI) algorithm \citep{bishop2006}, where each factor $q_j(\theta_j)$ is updated conditional on the others. The optimal update has the form
\begin{equation}
\label{eq:cavi_sofr}
\log q_j^*(\theta_j) = \E_{q(\boldsymbol{\theta}_{-j})} \!\big[\log p(\boldsymbol{y}, \boldsymbol{\theta})\big] + \text{constant},
\end{equation}
with $q(\boldsymbol{\theta}_{-j})$ denoting the variational distribution of all parameters except $\theta_j$, and ``constant'' refers to the terms independent of $\theta_j$.

\subsection{Variational EM}
\label{sec:vem}

Suppose that $\theta_l \in \Theta$ is a parameter whose full posterior distribution is intractable to obtain due to its non-conjugacy or is not of primary interest. VEM algorithms addresses this challenge by allowing such parameter to be inferred via maximization steps, while still estimating the posterior distributions for the remaining model parameters \citep{neal1998, dimitris2008}. 



Let $\Lambda \subset \Theta$ represent the set of model parameters for which we can approximate the posterior distribution using VB. The VEM algorithm iteratively maximizes the ELBO with respect to $q(\Lambda)$ and $\theta_l$ using the following two steps:

\begin{itemize}

\item \textbf{E-step}: Assuming $\theta_l$ fixed, we update the variational distribution $q^*(\Lambda)$ derived using the CAVI algorithm.

\item \textbf{M-step}: $\theta_l$ is updated by maximizing the ELBO, that is, $\theta_l = \arg\max_{\theta_l} \text{ELBO}(q^*(\Lambda), \theta_l)$, while holding the variational distribution $q^*(\Lambda)$ updated in the E-step fixed.

\end{itemize}

The VEM algorithm can be constructed in two ways. One approach, is to run the VB procedure in the  E-step until convergence before updating the remaining parameters in the M-step \citep{osborne2022,li2023}. This strategy takes longer to converge and it may be more sensitive to initialization of the algorithm. Alternatively, the VEM algorithm can be implemented by alternating a single E-step update with a single M-step update at each iteration. This strategy substantially reduces runtime and tends to be less sensitive to the algorithm's initialization \citep{liu2019, dacruz2024}. In this work, we employ the latter. In the next section, we present the VB equations corresponding to the E-step of our VEM algorithm. For the M-step, we maximize the ELBO with respect to the regularization parameters $\lambda^2_j$s to obtain their estimates via a closed-form solution. Our algorithm is summarized in Algorithm \ref{alg:vb_sofr}.

\subsection{VB update equations}
\label{sec:vb_sofr}

In this section, we derive the VB update equations for the variational distributions of the parameters from our proposed model in Equation \eqref{eq:hier_sofr}, except for regularization parameters $\boldsymbol{\lambda}^2 = (\lambda^2_1, \dotsc, \lambda^2_p)$, that is, in what follows, we derive the optimal variational distributions for $(\mathbf{b}, \boldsymbol{\theta}, \mathbf{Z}, \sigma^2, \boldsymbol{\tau}^2)$. The update of the regularization parameters is then computed by directly maximizing the ELBO with respect to each $\lambda^2_j$ given the obtained variational distributions. The VB updates rely on expectations taken with respect to the optimal variational distributions, whose derivations, along with the ELBO, are provided in Appendices \ref{sec:vb_exp_sofr} and \ref{sec:elbo_sofr}, respectively.

\paragraph{• Update equation for $q(\mathbf{b})$:}
Using the CAVI algorithm, we derive the update equation for $q(\mathbf{b})$, by computing the expectation of the logarithm of the complete-data likelihood with respect to all quantities except $\mathbf{b}$ as in Equation \eqref{eq:cavi_sofr}, where the complete-data likelihood is defined as follows:
\begin{align}
\label{eq:vb_complete_sofr}
    p(\mathbf{Y}, \mathbf{Z}, \mathbf{b}, \boldsymbol{\theta}, \boldsymbol{\tau}^2, \sigma^2, \boldsymbol{\lambda}^2)
    =&\ p(\mathbf{Y} \mid \mathbf{Z}, \mathbf{b}, \sigma^2) \times p(\mathbf{Z} \mid \boldsymbol{\theta}) \times p(\mathbf{b} \mid \boldsymbol{\tau}^2, \sigma^2)  \\
    & {} \times p(\boldsymbol{\theta})  \times p(\sigma^2) \times p(\boldsymbol{\tau}^2 \mid \boldsymbol{\lambda}^2).
    \nonumber
\end{align}

In what follows, we use $\upc$ to denote equality up to a constant. Since, only $p(\mathbf{Y} \mid \mathbf{Z}, \mathbf{b}, \sigma^2)$ and $p(\mathbf{b} \mid \boldsymbol{\tau}^2, \sigma^2)$ in Equation \eqref{eq:vb_complete_sofr} depend on $\mathbf{b}$, the update equation for $q(\mathbf{b})$ is given as
\begin{align}
\label{eq:vb_eb_sofr_1}
    \log q^*(\mathbf{b})
    &= \E_{q(-\mathbf{b})}[\log p(\mathbf{Y} \mid \mathbf{Z}, \mathbf{b}, \sigma^2)] + \E_{q(-\mathbf{b})}[\log p(\mathbf{b} \mid \boldsymbol{\tau}^2, \sigma^2)] +  \text{constant}
    \nonumber \\
    &\upc \E_{q(-\mathbf{b})}\left[ -\frac{(\boldsymbol{y} - W\Gamma\mathbf{b})\trp(\boldsymbol{y} - W\Gamma\mathbf{b})}{2\sigma^2}\right] + \E_{q(-\mathbf{b})}\left[-\sum_{k=1}^K\sum_{j=1}^p\frac{{b_{kj}}^2/\tau^2_{kj}}{2\sigma^2}\right].
\end{align}

By defining $\boldsymbol{\eta} = ({\boldsymbol{\eta}_1}, \dotsc, {\boldsymbol{\eta}_p})$, where ${\boldsymbol{\eta}_p} = \left(\frac{1}{\tau^2_{1p}}, \dotsc, \frac{1}{\tau^2_{Kp}}\right)\trp$. We can rewrite Equation \eqref{eq:vb_eb_sofr_1} as
\begin{align}
\label{eq:vb_eb_sofr_2}
    \log q^*(\mathbf{b})
    &= \E_{q(-\mathbf{b})}\left[-\frac{\mathbf{b}\trp\diag(\boldsymbol{\eta})\mathbf{b} + (\mathbf{y} - W\Gamma\mathbf{b})\trp(\mathbf{y} - W\Gamma\mathbf{b})}{2\sigma^2}\right] 
    \nonumber \\
    &\upc \E_{q(\sigma^2)}\left(\frac{1}{\sigma^2}\right)\left\{-\frac{\mathbf{b}\trp\left[\E_{q(\boldsymbol{\tau}^2)}\diag(\boldsymbol{\eta}) + \E_{q(\boldsymbol{Z})}(\Gamma W\trp W\Gamma)\right]\mathbf{b} - 2\mathbf{b}\trp \E_{q(\boldsymbol{Z})}(\Gamma)W\trp\mathbf{y}}{2}\right\}.
\end{align}

By completing the squares in Equation \eqref{eq:vb_eb_sofr_2}, we obtain
\begin{equation*}
\label{eq:vb_eb_sofr_3}
    \log q^*(\mathbf{b}) \upc -\frac{\left(\mathbf{b} - Q^{-1} \E_{q(\boldsymbol{Z})}(\Gamma)W\trp\mathbf{y}\right)\trp (\E_{q(\sigma^2)}(1/\sigma^2)Q)\left(\mathbf{b} - Q^{-1}\E_{q(\boldsymbol{Z})}(\Gamma)W\trp\mathbf{y}\right)}{2},
\end{equation*}
where $Q = \left(\E_{q(\boldsymbol{\tau}^2)}\diag(\boldsymbol{\eta}) +\E_{q(\boldsymbol{Z})}(\Gamma W\trp W \Gamma) \right)$. Thus, the optimal variational distribution of $\mathbf{b}$, $q^*(\mathbf{b})$ is a multivariate normal distribution with parameters
\begin{equation}
\label{eq:vb_b_1}
    \boldsymbol{\mu}_{\mathbf{b}} = Q^{-1}\E_{q(\boldsymbol{Z})}(\Gamma)W\trp\mathbf{y}
\end{equation}
and
\begin{equation}
\label{eq:vb_b_2}
    \Sigma_{\mathbf{b}} = (\E_{q(\sigma^2)}(1/\sigma^2)Q)^{-1}.
\end{equation}

\paragraph{• Update equation for $q(\theta_{j})$:}
The update equation for $q(\theta_{j})$ can be obtained similarly to $q(\mathbf{b})$, as follows:
\begin{align*}
\log q^*(\theta_j)
=&\ \E_{q(-\theta_j)}[\log p(\mathbf{Z} \mid \boldsymbol{\theta})] + \E_{q(-\theta_j)}[\log p(\boldsymbol{\theta})] + \text{constant}
\\
\upc&\ (0.5 - 1)\log \theta_j + (1 - 0.5 - 1)\log (1-\theta_j)\\
& {} + \E_{q(Z_j)}(Z_j)\log \theta_j + (1 - \E_{q(Z_j)}(Z_j)) \log(1-\theta_j)\\
=&\ \left(\E_{q(Z_j)}(Z_j) + 0.5 - 1\right)\log\theta_j + \left(2-\E_{q(Z_j)}(Z_j)-0.5-1\right)\log(1-\theta_j).
\end{align*}

$q^*(\theta_{j})$ is a Beta distribution with parameters
\begin{gather}
\label{eq:theta_sofr_1}
a_{j} = \E_{q(Z_j)}(Z_{j}) + 0.5
\\
\intertext{and}
\label{eq:theta_sofr_2}
b_{j} = 2 - \E_{q(Z_{j})}(Z_{j}) - 0.5.
\end{gather}

\paragraph{• Update equation for $q(Z_{j})$:}
\begin{equation}
\label{eq:qZj}
\log q^*(Z_j) = \E_{q(-Z_j)}(\log p(\mathbf{Y} \mid \mathbf{Z}, \mathbf{b}, \sigma^2)) + \E_{q(-Z_j)}(\log p(\mathbf{Z} \mid \boldsymbol{\theta})) + \text{constant}
\end{equation}

Taking the expectation in Equation \eqref{eq:qZj}, the terms that do not depend on $Z_j$ will be added to the constant term. Hence,
\begin{align*}
\log q^*(Z_j) \upc \sum_{r = 0}^1\mathrm{I}(Z_j = r)\Bigg\{ &-\frac{n}{2}\E_{q(\sigma^2)}(\log(\sigma^2)) \\
&-\frac{1}{2}\E_{q(\sigma^2)}\left(\frac{1}{\sigma^2}\right)\E_{q(\mathbf{Z}_{-j})q(\mathbf{b})}(\mathbf{y} - W\tilde{\Gamma}\mathbf{b})\trp(\mathbf{y} - W\tilde{\Gamma}\mathbf{b})\\
& + r\E_{q(\theta_j)}(\log(\theta_j)) + (1 - r)\E_{q(\theta_j)}(\log(1 - \theta_j))\Bigg\},
\end{align*}
where $\tilde{\Gamma} = \diag{\tilde{\boldsymbol{\gamma}}}$, $\tilde{\boldsymbol{\gamma}} =(Z_1, \dotsc, Z_{j-1}, r, Z_{j+1},\dotsc, Z_p)\trp\otimes\boldsymbol{1}_K$.

Thus,
$q^*(Z_j)$ is a Bernoulli distribution with probability of success $pz_{j}$ given as
\begin{equation}
\label{eq:vb_zj_sofr_3}
pz_{j} = \frac{\exp(u_{j1})}{\sum_{r = 0}^1\exp(u_{jr})},
\end{equation}
where
\begin{align*}
   u_{jr} = &-\frac{n}{2}\E_{q(\sigma^2)}(\log(\sigma^2)) \\
&-\frac{1}{2}\E_{q(\sigma^2)}\left(\frac{1}{\sigma^2}\right)\E_{q(\mathbf{Z}_{-j})q(\mathbf{b})}(\mathbf{y} - W\tilde{\Gamma}\mathbf{b})\trp(\mathbf{y} - W\tilde{\Gamma}\mathbf{b}) \\
& + r\E_{q(\theta_j)}(\log(\theta_j)) + (1 - r)\E_{q(\theta_j)}(\log(1 - \theta_j)).
\end{align*}

\paragraph{• Update equation for $q(\sigma^2)$:}
\begin{align*}
\log q(\sigma^2)
=&\ \E_{q(-\sigma^2)}[\log p(\mathbf{y} \mid \mathbf{b}, \mathbf{Z}, \sigma^2)] + \E_{q(-\sigma^2)}[\log p( \mathbf{b} \mid \boldsymbol{\tau}^2, \sigma^2)]  + \E_{q(-\sigma^2)}[\log p(\sigma^2)] + \text{constant}\\
\upc&\ \E_{q(-\sigma^2)}\Bigg\{\frac{n}{2}\log(1/\sigma^2) + \frac{Kp}{2}\log(1/\sigma^2) - \frac{1}{2\sigma^2}(\mathbf{y} - W\Gamma\mathbf{b})\trp(\mathbf{y} - W\Gamma\mathbf{b})\\
& {} - \frac{1}{2\sigma^2}\sum_{j=1}^p\sum_{k = 1}^K\left(\frac{b_{kj}^2}{\tau^2_{kj}}\right) + (\delta_1 +1)\log(1/\sigma^2) - \delta_2/\sigma^2\Bigg\}
\\
=&\ \left(\frac{n + Kp}{2} + \delta_1 + 1\right)\log(1/\sigma^2)\\
& {} - \frac{1}{\sigma^2}\left(\frac{\E_{q(\mathbf{Z})q(\mathbf{b})}(\mathbf{y} - W\Gamma\mathbf{b})\trp(\mathbf{y} - W\Gamma\mathbf{b}) + \sum_{j=1}^p\sum_{k = 1}^K\E_{q(\boldsymbol{\tau}^2)}\left(\frac{1}{\tau^2_{kj}}\right)\E_{q(\boldsymbol{b})}\left(b_{kj}^2\right)}{2} + \delta_2\right).
\end{align*}

$q^*(\sigma^2)$ is an inverse-gamma distribution with parameters
\begin{equation*}
\delta^*_1 = \frac{n + Kp + 2\delta_1}{2}
\end{equation*}
and
\begin{equation}
\label{eq:sigma2_sofr_2}
\delta^*_2 = \frac{1}{2}\left(\E_{q(\mathbf{Z})q(\mathbf{b})}(\mathbf{y} - W\Gamma\mathbf{b})\trp(\mathbf{y} - W\Gamma\mathbf{b}) + \sum_{j = 1}^{p}  \sum_{k = 1}^{K} \left[\E_{q(\boldsymbol{\tau}^2)}\left(\frac{1}{\tau^2_{jk}}\right) \times \E_{q(\mathbf{b})} (b_{kj}^2)\right] + 2\delta_2\right).
\end{equation}

\paragraph{• Update equation for $q(\tau_{kj}^2)$:}
\begin{align*}
    \log q({\tau^2_{kj}})
    &= \E_{q(-\tau^2_{kj})}[\log p(b_{kj} | \sigma^2, \tau^2_{kj})] + \E_{q(-\tau^2_{kj})}[\log p(\tau^2_{kj} \mid \lambda^2_j)] + \text{constant}\\
    &\upc \E_{q(-\tau^2_{kj})}\left\{ \frac{1}{2}\log\left(\frac{1}{\tau^2_{kj}} + \frac{1}{\sigma^2}\right) - \left(\frac{b^2_{kj}}{2}\times\frac{1}{\sigma^2}\times\frac{1}{\tau^2_{kj}} + \frac{\lambda^2_j}{2}\tau^2_{kj}\right)\right\}
    \\
    &\upc -\frac{1}{2}\log(\tau^2_{kj}) - \frac{1}{2}\left(\left[\E_{q(\mathbf{b})}(b^2_{kj})\E_{q(\sigma^2)}\left(\frac{1}{\sigma^2}\right)\right]\times\frac{1}{\tau^2_{kj}} + \lambda^2_j\tau^2_{kj}\right).
\end{align*}

$q^*(\tau_{kj}^2)$ is a generalized-inverse-Gaussian (GIG) distribution with parameters
\begin{equation}
\label{eq:tau2_1}
p = 1/2, \;\; \chi_{kj} = \E_{q(\mathbf{b})} (b_{kj}^2) \times \E_{q(\sigma^2)}\left(\frac{1}{\sigma^2}\right) \text{and} \;\; \psi_{kj} = \lambda^2_j.
\end{equation}


\begin{algorithm}
\caption{The variational EM algorithm for variable selection in scalar-on-function regression.}
\label{alg:vb_sofr}
\begin{algorithmic}[1]
\State Set hyperparameter values for the prior distributions;
\State Assign initial values to $pz_1, \dotsc, pz_p, \delta_2^*, \chi_{kj}$, $j = 1, \dotsc, p$, $k = 1, \dotsc, K$, and $\boldsymbol{\lambda}^2$;
\While{$\mathrm{ELBO}^{(c)} - \mathrm{ELBO}^{(c - 1)} > \text{tolerance}$ and $c \leq N_{\mathrm{iter}}$}
\State Update the parameters of $q^*(\mathbf{b})$ using Equations \eqref{eq:vb_b_1} and \eqref{eq:vb_b_2}
\State Update the parameters of $q^*(\sigma^2)$ using Equation \eqref{eq:sigma2_sofr_2};
    \For{$j = 1, \dotsc, p$}
        \For{$k = 1, \dotsc, K$}
            \State Update the parameters of $q^*(\tau_{kj}^2)$ using Equation \eqref{eq:tau2_1};
        \EndFor
    \EndFor
    \For{$j = 1, \dotsc, p$}
            \State Update the parameters of $q^*(\theta_{j})$ using Equations \eqref{eq:theta_sofr_1} and \eqref{eq:theta_sofr_2};
            \State Update the parameter of $q^*(Z_{j})$ using Equation \eqref{eq:vb_zj_sofr_3}.
    \EndFor
    \State Update $\lambda^2_j$, $j = 1, \dotsc, p$ as follows
    \begin{equation*}
        {\lambda^2_j}^{(c)} = \frac{2K}{\sum_{k = 1}^KE_{q^*(\tau_{kj}^2)}(\tau^2_{kj})};
    \end{equation*}
    \State Calculate the current $\text{ELBO}$, $\text{ELBO}^{(c)}$ as derived in Appendix \ref{sec:elbo_sofr}.
  \EndWhile
\end{algorithmic}
\end{algorithm}

\clearpage
\section{Simulation studies}

In this section, we present three simulation studies. The first two focus on scenarios involving only functional covariates, while the third addresses a partially functional regression model. For each study, we assess the performance of our proposed method under different conditions by varying the error variance $\sigma^2$, the number of observations $n$, or both. We also compare the performance of our method against existing approaches used for variable selection in linear regression, more specifically, we compare our approach to  grLASSO (group least absolute shrinkage and selection operator) \citep{yuan2006}, grMCP (group minimax concave penalty) \citep{huang2012}, grSCAD (group smoothly clipped absolute deviation) \citep{wang2007} and BGLSS (Bayesian group LASSO with spike-and-slab priors) \citep{ghosh2015}. All integrals in the simulation studies are computed using the trapezoidal rule.

\subsection{Simulation Design}
\label{sec:sim_sofr}

\subsubsection{Simulation Study 1}

In Simulation Study 1, we focus on the model presented in Equation \eqref{eq_sofr_model_vs}, where we consider only functional covariates. We evaluate the performance of our proposed method in recovering the relevant covariates and in estimating the final functional coefficients, when the partial functional coefficients are generated as linear combinations of basis coefficients sampled according to our hierarchical Bayesian framework described in Equation \eqref{eq:hier_sofr}. We consider two functional covariates, each observed at 100 equally-spaced points on $[0, 1]$, with only the first covariate being related to the scalar response. The functional covariates are generated using four cubic B-splines basis functions given as $X_{ij}(t) = \sum_{k=1}^{4}a_{ijk}B_{jk}(t),$ $t \in [0, 1]$, $i = 1, \dotsc, n$ and $j = 1, 2$. The basis coefficients for $X_{ij}(\cdot)$ are generated as $a_{ijk}  = a_{jk} + e_{ijk}$, where $e_{ijk} \sim N(0, 100)$, $a_{1k} \sim N(5, 100)$ and $a_{2k} \sim N(2, 1)$.

For computational simplicity, we assume the same four basis functions used to construct the functional covariates to obtain the partial functional coefficients $\beta_{j}(t),$ $j = 1, 2$. The basis coefficients for the relevant covariate $b_{1k},$ for $k = 1, \dotsc, 4$, are drawn from our hierarchical Bayesian model, with priors specified as univariate normal distributions with a mean of zero and a variance of $\sigma^2\tau^2_{1k}$. The local regularization parameters $\tau^2_{1k},$ for $k = 1, \dotsc, 4$, are sampled from their prior, with $\lambda^2_1$ set to $0.001$ to ensure that the basis coefficients for the relevant covariate are significantly different than zero. The basis coefficients for the non-relevant covariate are set to zero, that is, $b_{2k} = 0,$ for all $k$.  Finally, the scalar response is generated according to the model in Equation \eqref{eq_sofr_model_vs}, with an intercept of $\beta_0 = 10$.

We simulate data under six scenarios by varying both the error variance ($\sigma^2$ = 0.1 and 0.5) and the sample size ($n \in \{50, 100, 200\}$). Note that since $\sigma^2$ appears in the prior variance of the basis coefficients, varying $\sigma^2$ alters the scale of $\beta_1(\cdot)$ but not its shape. For each combination of variance and sample size, we generate $S = 100$ datasets to evaluate the performance of our method.

\subsubsection{Simulation Study 2}

We compare the performance of our proposed method with grLASSO, grMCP, grSCAD and BGLSS, linear regression approaches that have been adapted and used for variable selection in functional regression \citep{matsui2011, collazos2016}.
 In this setting, we generate four functional covariates, each observed at 81 equally-spaced points on $[0, 1]$. Covariates 1 and 3 are relevant to the response, whereas Covariates 2 and 4 are not associated with the scalar outcome. Each  covariate is generated as $X(t) = 5\sum_{k = 1}^{10}c_k\gamma_k(t),$ where $c_k \sim N(0, k^{-2})$, $\gamma_1(t) = 1$ and $\gamma_{k+1}(t) = \sqrt{2}\cos(k\pi t)$ for $k = 1, \dotsc, 10$. A similar way of generating the functional covariates was adopted in \citet{mbina2025}. The true partial functional coefficients are taken to be $\beta_1(t) = 2\sin(\pi t)$, $\beta_3(t) = 1.25\sin(3\pi t),$ and $\beta_2(t) = \beta_4(t) = 0,$ for all $t \in [0,1]$. The scalar response is generated as in Equation \eqref{eq_sofr_model_vs}, with an intercept of $\beta_0 = 20$.

Similar to Simulation Study 1, we generate $S = 100$ datasets for each scenario,  defined by a combination of variance and sample size. In this setting, we consider $\sigma^2 = 0.01$ and $0.05$ and $n \in \{100, 400\}$.

\subsubsection{Simulation Study 3}

In Simulation Study 3, we consider a partially functional regression model, focusing on variable selection of both scalar and functional covariates. The model is defined as follows:
\begin{equation}
\label{sofr_mixed}
    Y_i = \beta_0 + \sum_{j=1}^{p}\int_{0}^{1} X_{ij}(t)Z_j\beta_j(t)dt + \sum_{l=1}^{q} X_{il}^s u_{l} \alpha_l + \varepsilon_i, \; i = 1\dots, n, \;\;\; t \in [0,1],
\end{equation}
where $q$ is the number of scalar covariates, $X_{il}^s$ represents the $l$th scalar covariate for the $i$th observation, $u_l$ is the latent indicator for whether $X_{il}^s$ is relevant, and $\alpha_l$ denotes the effect of the $l$th scalar covariate on $Y$. All other quantities are defined as in Section \ref{sec:model_sofr}.

In this simulation study, we generate two functional covariates, $X_j(t)$ for $j = 1, 2$, along with their associated partial functional coefficients, using the procedure described in Simulation Study 1, with the only change being that we consider six B-splines basis functions. In addition, we generate two scalar covariates, $X_1^s$ and $X_2^s$, sampled independently from normal distributions with means of $10$ and $20$, respectively, and variances of $4$. Only $X_1(t)$ and $X_2^s$ are considered relevant, meaning only the first functional covariate and the second scalar covariate should be retained in the model. The scalar coefficient associated to the relevant scalar covariate $X_2^s$, is sampled from its prior univariate normal distributions with a mean of zero and a variance of $\nu_2^2\sigma^2$, while ensuring that the sampled coefficient is sufficiently different than zero. The scalar coefficient associated with the non-relevant scalar covariates is set to zero, that is $\alpha_1 = 0$. The scalar response is then generated according to Equation \eqref{sofr_mixed}, with an intercept of $\beta_0 = 30$.

We simulate data under four scenarios by varying the sample size $n \in \{50, 100\}$ and error variance $\sigma^2$ (0.1 and 0.5). For each scenario, we generate $S = 100$ datasets to evaluate the performance of our method.

\subsection{Initialization and performance metrics}
\label{sec:sofr_ini}

We adopt a weakly informative prior setting with $\delta_1 = 0.01$ and $\delta_2 = 0.01$, indicating no strong prior knowledge about the error variance.
For the initialization of the VEM algorithm, we set the parameter $\delta_2^*$ in the variational distribution of the error variance so that the mean of the variational distribution of $\sigma^2$ matches the true error variance used in generating the data, while maintaining a relatively large variance in the distribution. For the regularization parameters $\lambda^2_j$s, we set their initial values arbitrarily. In the simulation studies, all inclusion probabilities $pz{_j}$ in Equation \eqref{eq:vb_zj_sofr_3} are initially set to one, reflecting the assumption that all covariates are potentially relevant. Alternatively, multiple random initializations may be employed and selecting the configuration showing the highest ELBO value, as implemented in the real data analysis. The algorithm runs with an ELBO convergence threshold of 0.01 and a maximum of 100 iterations.

To evaluate the performance of the methods in estimating the functional coefficients, we compute the empirical mean integrated squared error (EMISE) for each final functional coefficient across simulated datasets. Since in our simulation studies we generate functional coefficients with an equal number of evaluation points $n_t$. The EMISE is obtained as follows:

\begin{equation}
\label{eq:EMISE}
\mathrm{EMISE}_j = \frac{1}{S} \sum_{s=1}^S \left[ \frac{T}{n_t}\sum_{m = 1}^{n_t} \left(\xi_j(t_m) - \hat{\xi}^s_j(t_m)\right)^2 \right],
\end{equation}
where $s$ indicates the $s$th simulated dataset, $T$ represents the interval length in which the functional coefficients are evaluated with $n_t$ equally-spaced observed evaluation points, $\xi_j(t_m)$ is the true value of the $j$th final functional coefficient evaluated at point $t_m$ and $\hat{\xi}^s_j(t_m)$ is its corresponding estimated value for the $s$th simulated dataset. For our proposed method, 
\begin{equation*}
    \hat{\xi}_j^{\,s}(t_m) = \hat{Z}_j^{\,s}\hat{\beta}_j^{\,s}(t_m),
\end{equation*}
where $\hat{\beta}_j^{s}(t_m)$ is calculated as $\mathbf{B}_j(t_m)^\top \boldsymbol{\mu}_{b_j}^{s}$, and $\hat{Z}_j^{\,s}$ is the mode of the variational distribution of $Z_j$, and $\boldsymbol{\mu}_{b_j}^{\,s}$ is the mean of the variational distribution of $\mathbf{b}_j$ for the $s$th dataset. For the other methods compared in Simulation Study 2, $\xi_j(t)$ corresponds to $\beta_j(t)$ in Equation \eqref{eq_sofr_model_vs}, and $\hat{\xi}_j^{\,s}(t_m)$ to $\hat{\beta}_j^{\,s}(t_m)$, since non-relevant functional coefficients are directly estimated as zero.

Additionally, we evaluate the goodness of fit of each method using the mean squared error (MSE), computed by comparing the predicted and observed responses for each simulated dataset and each scenario. To assess variable selection performance, we report the proportion of times each covariate is selected across the 100 simulated datasets in each scenario. For the VEM method, a covariate is included in the model when the mode of the variational distribution of its corresponding $Z_j$ is one, which is equivalent to using a threshold of 0.5 on its inclusion probability $pz_j$.

\subsection{Simulation results}
\label{sec:sim_res_sofr}

In this section, we present the numerical results for all simulation studies described in Section \ref{sec:sim_sofr}. For Simulation Studies 1 and 3, where the functional covariates and their corresponding coefficients are generated as linear combinations of B-splines basis functions, we fit our model using the same set of B-splines employed in the data generation for simplicity. In Simulation Study 2, we use the set with seven B-splines to compare the performance of our method with grLASSO, grMCP, grSCAD and BGLSS. The number of basis functions $K$, was chosen based on the visual inspection of the mean estimated curves for the functional coefficients. Alternatively, one could select the optimal number of basis functions via generalized cross-validation, as is implemented in the real data analyses.

For all simulation studies, we standardize the functional covariates prior to model fitting by subtracting each one by its average, and dividing it by its standard deviation, as significant differences in scale can affect shrinkage and potentially reduce prediction accuracy. The average value for the $j$th functional covariate at each evaluation point $t_m$ is computed as $\bar{X}_j(t_m) = \frac{\sum_{i = 1}^{n}X_{ij}(t_m)}{n}$, where $m \in 1, \dotsc, n_t$, with $n_t$ representing the number of evaluation points for the functional covariates, assumed to be either $100$ in Simulation Studies 1 and 3, or $81$ in Simulation Study 2. The standard deviation of the $j$th functional covariate at point $t$ is given as $\mathrm{SD}_j(t_m) = \frac{\sum_{i = 1}^{n}(X_{ij}(t_m)-\bar{X}_j(t_m))^2}{n - 1}$. We also center the response variable by subtracting it by its average value. After fitting the model, estimated coefficients and predicted responses are rescaled back, so that the results are presented in the original scale.

\subsubsection{Simulation Study 1}

In Table \ref{tab:sim1est_sofr}, we present the goodness-of-fit results for Simulation Study 1 across all scenarios using the MSE metric averaged over the 100 simulated datasets. To evaluate accuracy of our method in estimating the functional coefficients we report the EMISE metric for each final functional coefficient in each scenario. Overall, the quality of fit improves as the number of observations increases and the variance decreases, as evidenced by the decreasing average MSE values. As expected, the estimation of the final functional coefficients, particularly for the non-zero functional coefficient $\xi_1(\cdot)$, improves with larger sample sizes and lower error variances, as shown by the decreasing EMISE values.

\begin{table}[ht]
\centering
\caption[Simulation Study 1. MSE and EMISE values.]{Simulation Study 1. Mean squared error (MSE) values to assess goodness of fit taken as the average across the 100 simulated datasets for each scenario evaluated at different error variances ($\sigma^2 = 0.1, 0.5$) and sample sizes ($n = 50, 100, 200$). The empirical mean integrated squared error (EMISE) is used to evaluate the estimation of the final functional coefficients and is computed as in Equation \eqref{eq:EMISE}.}
\label{tab:sim1est_sofr}
\begin{tabular}{lccc ccc}
\toprule
 & \multicolumn{3}{c}{$\sigma^2 = 0.1$} & \multicolumn{3}{c}{$\sigma^2 = 0.5$} \\
\cmidrule(lr){2-4} \cmidrule(lr){5-7}
 & $n=50$ & $n=100$ & $n=200$ & $n=50$ & $n=100$ & $n=200$ \\
\midrule
  MSE & 0.1268 & 0.1165 & 0.1149 & 0.6355 & 0.6306 & 0.5688 \\ 
  EMISE$_{_1}$ & 0.0470 & 0.0278 & 0.0251 & 0.3121 & 0.1786 & 0.1273 \\ 
  EMISE$_{_2}$ & 0.0014 & 0.0002 & 0.0001 & 0.0067 & 0.0030 & 0.0005 \\ 
\bottomrule
\end{tabular}
\end{table}

Table \ref{tab:sim1z_sofr} reports the proportion of times each covariate is selected across the 100 simulated datasets. As expected, Covariate 1, which is the only truly relevant covariate, is always selected across all scenarios, while the accuracy in excluding the non-relevant covariate (Covariate 2) improves with larger sample sizes. 

\begin{table}[ht]
\centering
\caption[Simulation Study 1. Covariate selection performance.]{Simulation Study 1. Proportion of datasets each covariate is selected across the 100 simulated datasets for each scenario evaluated at different error variances ($\sigma^2 = 0.1, 0.5$) and sample sizes ($n = 50, 100, 200$). Variable selection is conducted as described in Section \ref{sec:sofr_ini}.}
\label{tab:sim1z_sofr}
\begin{tabular}{lrrrrrr}
\toprule
&  \multicolumn{3}{c}{$\sigma^2 = 0.1$} & \multicolumn{3}{c}{$\sigma^2 = 0.5$} \\
\cmidrule(lr){2-4} \cmidrule(lr){5-7}
  & $n = 50$ & $n = 100$ & $n = 200$ & $n = 50$ & $n = 100$ & $n = 200$ \\
\midrule
Covariate 1 & 1.00 & 1.00 & 1.00 & 1.00 & 1.00 & 1.00 \\ 
Covariate 2 & 0.13 & 0.06 & 0.01 & 0.20 & 0.12 & 0.04 \\ 
\bottomrule
\end{tabular}
\end{table}

\begin{figure}[ht]
    \centering
    \begin{subfigure}[b]{\textwidth}
        \centering
        \includegraphics[width=\textwidth]{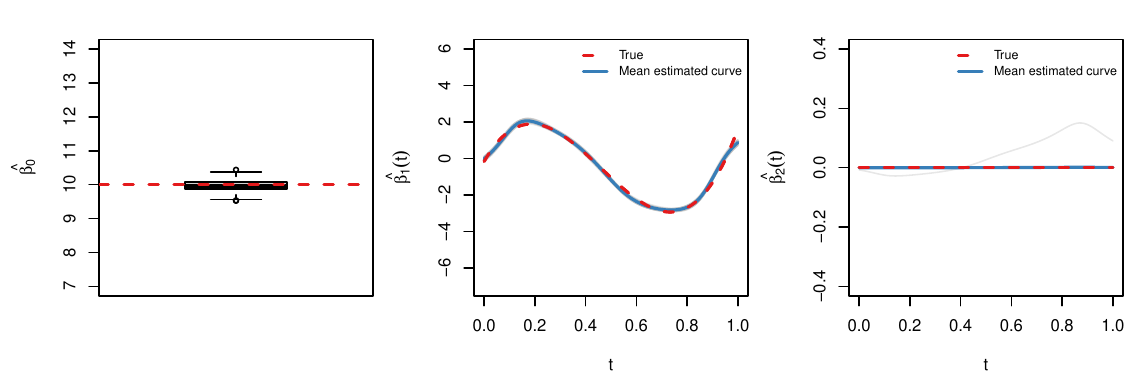}
        \caption{$\sigma^2 = 0.1$}
    \end{subfigure}
    \hfill
    \begin{subfigure}[b]{\textwidth}
        \centering
        \includegraphics[width=\textwidth]{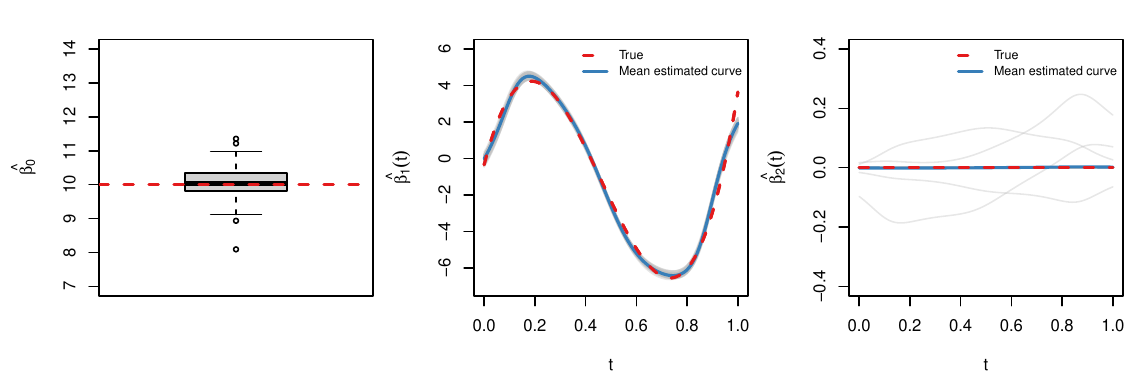}
        \caption{$\sigma^2 = 0.5$}
    \end{subfigure}

    \caption[Simulation Study 1. Estimated partial functional coefficients per dataset and boxplot of the intercept when $n = 200$.]{Simulation Study 1. Mean estimated curves for the two partial functional coefficients (blue), individual estimates per simulated dataset (grey), and boxplots of the estimated intercept across 100 simulated datasets, for sample size $n = 200$ and varying error variance ($\sigma^2 = 0.1, 0.5$). True values for the partial functional coefficients and intercept are shown in red.}
    \label{fig:sim1_all_sofr}
\end{figure}

Figure \ref{fig:sim1_all_sofr} presents the mean estimated curves for both partial functional coefficients, individual estimates from each simulated dataset, and boxplots of the estimated intercept across 100 simulations for scenarios with sample size $n = 200$ (results for the other scenarios are provided in Section 2 of the Supplementary Material). The mean estimated curve for each partial functional coefficient are computed as $\hat\beta_j(t_m) = \frac{1}{100}\sum_{s = 1}^{100}\mathbf{B}_{j}(t_m)\trp \mathbf{\mu_b}^s_{j},$ for each evaluation point $t_m$, where $\mathbf{B}_{j}(t_m)$ corresponds to the vector of four B-splines basis functions computed at the evaluation point $t_m$ and is fixed across all the $S = 100$ simulated datasets, and ${\mu_b}^s_j$ refers to the estimated mean of the variational distribution of $\mathbf{b}_j$ for the $s$th simulated dataset. Across all scenarios, the mean estimated curves closely align with the true partial functional coefficients, and the estimated intercept values also align with the true value, as indicated by the median being near 10. Note that, although some estimated curves (in grey) for $\beta_2(\cdot)$ differ from zero, they are on a small scale close to zero, resulting in small EMISE values in Table \ref{tab:sim1est_sofr}. Overall, estimation accuracy improves with larger sample sizes for both intercept and functional coefficients as shown by the decreasing variance in the boxplots and by the estimated individual curves represented in grey, as expected. Note that, by varying the error variance in the scenarios, the scale of the non-zero  functional coefficient increases per simulation design, but its shape remains the same, allowing us to still be able to compare the results between the two scenarios.

\begin{figure}[ht]
    \centering
    \begin{subfigure}[b]{0.48\textwidth}
        \centering
        \includegraphics[width=\textwidth, page = 1]{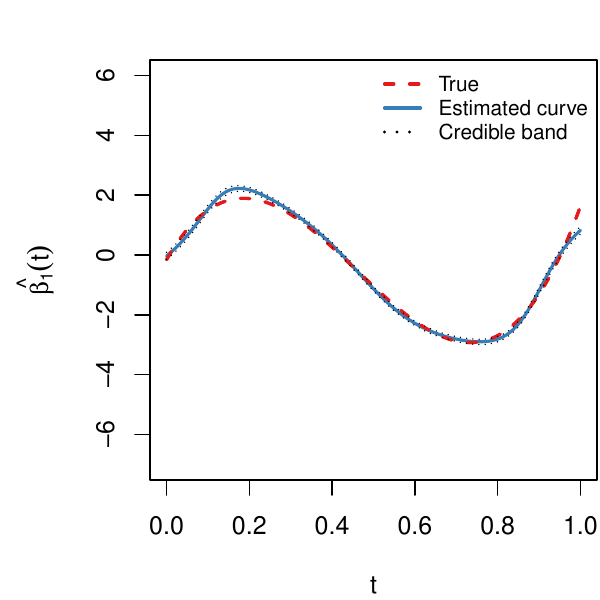}
        \caption{$\sigma^2 = 0.1$}
    \end{subfigure}
    \hfill
    \begin{subfigure}[b]{0.48\textwidth}
        \centering
        \includegraphics[width=\textwidth, page = 1]{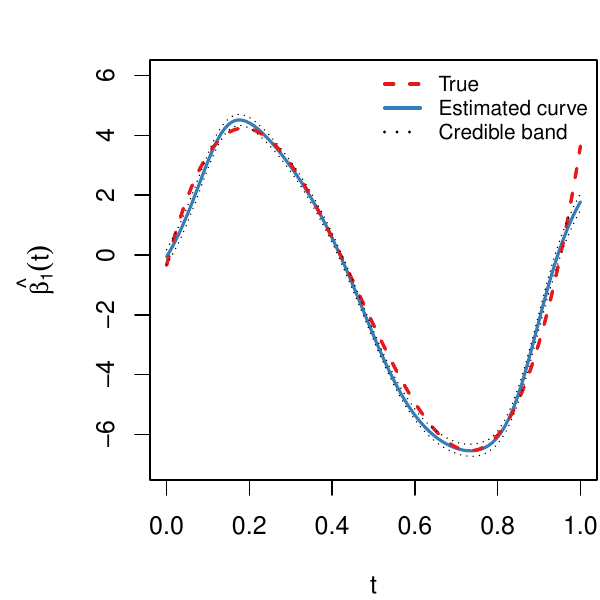}
        \caption{$\sigma^2 = 0.5$}
    \end{subfigure}
    \caption[Simulation Study 1. Estimated non-zero partial functional coefficient with credible bands when $n = 200$.]{Simulation Study 1. Estimated curves for the non-zero partial functional coefficient (blue) with 95\% credible bands (black) for one of the simulated datasets under the scenario with sample size $n = 200$ and error variances $\sigma^2 = 0.1, 0.5$. The true values for the partial functional coefficients are shown in red.}
    \label{fig:sim1_cred_sofr}
\end{figure}

As an illustration of the construction of credible bands for the functional coefficients, Figure \ref{fig:sim1_cred_sofr} displays the estimated non-zero partial functional coefficient along with 95\% credible bands for one of the simulated datasets under scenarios with sample size $n = 200$ (results for the other scenarios are provided in Section 2 of the Supplementary Material). The estimated partial functional coefficient associated to the non-relevant covariate is zero over the entire domain across all scenarios. We construct the credible bands by adapting the approach of \citet{dacruz2024} to the context of SoFR. Specifically, we draw 200 samples from the variational distributions of each $Z_j$ and the corresponding vector of basis coefficients $\mathbf{b}_j$. For each sample, we compute $Z_j\mathbf{b}_j$, shrinking the basis coefficients of the $j$th functional coefficient to zero whenever $Z_j = 0$. Then, we generate 200 curves, evaluated at a common grid of points. At each evaluation point, the 2.5\% and 97.5\% quantiles of the sampled function values are used to construct the 95\% credible band. Notably, the 95\% credible bands cover most of true partial functional coefficient $\beta_1(\cdot)$ for most scenarios (with the exception of $n=50$, shown in Figure 3 in Section 2 of the Supplementary Material). 


\subsubsection{Simulation Study 2}

In Simulation Study 2, we obtain estimates of the functional coefficients for grLASSO, grMCP, and grSCAD using the \fn{cv.grpreg} function from the \R Package \pck{grpreg} \citep{breheny2009, breheny2015}, which automatically selects the regularization parameter via cross-validation, a convenient feature given that we fit 100 simulated datasets. For BGLSS, estimates are obtained using the \fn{BGLSS} function from the \R package \pck{MBSGS} \citep{liquet2017}. We ran the Gibbs sampler implemented in the package for 10,000 iterations, with the a burn-in of 5000 samples.

In Table \ref{tab:sim2est_sofr}, we compare the overall performance of all approaches based on the MSE and EMISE metrics. For each scenario, we report the average MSE across the 100 simulated datasets. All methods demonstrate satisfactory goodness of fit, with our VEM algorithm and BGLSS showing the smallest average MSE values, indicating higher predictive performance. Comparing the methods with respect to the accuracy in estimating the final functional coefficients based on the EMISE values, once again our VEM algorithm and BGLSS show the best performances across all scenarios, with perfect estimation of the final functional coefficients that should not be in the model (Covariates 2 and 4). The other approaches have difficulties in accurately estimating the final functional coefficients, specially for the relevant covariates (Covariates 1 and 3) as evidenced by the high EMISE values across all scenarios. Overall, across all methods, as expected, the estimation of the final functional coefficients improves as the error variance decreases and the sample size increases.

\begin{table}[ht]
\centering
\caption[Simulation Study 2. MSE and EMISE values.]{Simulation Study 2. Mean squared error (MSE) values to assess goodness of fit of each method taken as the average across the 100 simulated datasets for each scenario evaluated at different error variances ($\sigma^2 = 0.01, 0.05$) and sample sizes ($n = 100, 400$). The empirical mean integrated squared error (EMISE) is used to evaluate the estimation of the final functional coefficients, as computed in Equation \eqref{eq:EMISE}}
\label{tab:sim2est_sofr}
\begin{tabular}{*{3}{l}*{5}{r}}
\toprule
$n$ & $\sigma^2$ & Metric & VEM & BGLSS & grLASSO & grMCP & grSCAD \\
\midrule
   &  & MSE & 0.0087 & 0.0091 & 0.0119 & 0.0115 & 0.0114 \\ 
   &  & EMISE$_{_1}$ & 0.0147 & 0.0138 & 96.3575 & 95.9480 & 96.6311 \\ 
   & $0.01$ & EMISE$_{_2}$ & 0.0000 & 0.0000 & 26.5215 & 4.7412 & 2.9211 \\ 
   &  & EMISE$_{_3}$ & 0.0345 & 0.0165 & 114.9012 & 106.7503 & 107.4601 \\ 
  100 &  & EMISE$_{_4}$ & 0.0000 & 0.0000 & 22.6902 & 3.2808 & 1.4446 \\ 
   \cmidrule{2-8} &  & MSE & 0.0435 & 0.0443 & 0.0973 & 0.0847 & 0.0832 \\ 
   &  & EMISE$_{_1}$ & 0.0395 & 0.0357 & 836.4752 & 793.1099 & 785.3696 \\ 
   & $0.05$ & EMISE$_{_2}$ & 0.0000 & 0.0000 & 269.2930 & 43.4807 & 23.4547 \\ 
   &  & EMISE$_{_3}$ & 0.0706 & 0.0333 & 719.9380 & 649.9282 & 641.3769 \\ 
   &  & EMISE$_{_4}$ & 0.0000 & 0.0000 & 261.7394 & 53.8989 & 40.8945 \\ 
   \midrule &  & MSE & 0.0098 & 0.0099 & 0.0109 & 0.0107 & 0.0107 \\ 
   &  & EMISE$_{_1}$ & 0.0073 & 0.0060 & 26.1627 & 24.8795 & 25.0139 \\ 
   & $0.01$ & EMISE$_{_2}$ & 0.0000 & 0.0000 & 7.9148 & 1.2193 & 0.8960 \\ 
   &  & EMISE$_{_3}$ & 0.0114 & 0.0068 & 25.1809 & 24.4853 & 24.4720 \\ 
   &  & EMISE$_{_4}$ & 0.0000 & 0.0000 & 7.3513 & 1.0777 & 1.0400 \\ 
   \cmidrule{2-8}400 &  & MSE & 0.0486 & 0.0490 & 0.0550 & 0.0539 & 0.0539 \\ 
   &  & EMISE$_{_1}$ & 0.0175 & 0.0145 & 120.5148 & 120.2662 & 121.0753 \\ 
   & $0.05$ & EMISE$_{_2}$ & $5.50 \times 10^{-7}$ & 0.0000 & 39.8264 & 2.1589 & 4.4621 \\ 
   &  & EMISE$_{_3}$ & 0.0360 & 0.0169 & 127.9857 & 126.6712 & 127.4303 \\ 
   &  & EMISE$_{_4}$ & 0.0000 & 0.0000 & 44.8066 & 8.6796 & 7.0458 \\ 
\bottomrule
    \end{tabular}
\end{table}

\begin{table}[ht]
\centering
\caption[Simulation Study 2. Covariate selection performance.]{Simulation Study 2. Proportion of datasets each covariate is selected across the 100 simulated datasets for each method and scenario evaluated at different error variances ($\sigma^2 = 0.01, 0.05$) and sample sizes ($n = 100, 400$).}
\label{tab:sim2_z_sofr}
\begin{tabular}{*{3}{c}*{5}{r}}
\toprule
$n$ & $\sigma^2$ & Covariate & VEM & BGLSS & grLASSO & grMCP & grSCAD \\
\midrule
\multirow{8}{*}{100} & \multirow{4}{*}{$0.01$}
  &  1 & 1.00 &     1.00    & 1.00 &  1.00   & 1.00 \\
& &  2 & 0.00 &     0.00    & 0.90 &  0.06   & 0.11 \\
& &  3 & 1.00 &     1.00    & 1.00 &  1.00   & 1.00 \\
& &  4 & 0.00 &     0.00    & 0.87 &  0.07   & 0.09 \\
\cmidrule(l){2-8}
& \multirow{4}{*}{$0.05$}
  & 1 & 1.00 &     1.00    & 1.00 &  1.00   & 1.00 \\
& & 2 & 0.00 &     0.00    & 0.96 &  0.08   & 0.21 \\
& & 3 & 1.00 &     1.00    & 1.00 &  1.00   & 1.00 \\
& & 4 & 0.00 &     0.00    & 0.90 &  0.10   & 0.19 \\
\midrule
\multirow{8}{*}{$400$} & \multirow{4}{*}{$0.01$}
  &  1 & 1.00 &     1.00    & 1.00 &  1.00   & 1.00 \\
& &  2 & 0.00 &     0.00    & 0.89 &  0.08   & 0.11 \\
& &  3 & 1.00 &     1.00    & 1.00 &  1.00   & 1.00 \\
& &  4 & 0.00 &     0.00    & 0.86 &  0.07   & 0.16 \\
\cmidrule(l){2-8}
& \multirow{4}{*}{$0.05$}
  & 1 & 1.00 &     1.00    & 1.00 &  1.00   & 1.00 \\
& & 2 & 0.01 &     0.00    & 0.86 &  0.13   & 0.20 \\
& & 3 & 1.00 &     1.00    & 1.00 &  1.00   & 1.00 \\
& & 4 & 0.00 &     0.00    & 0.90 &  0.11   & 0.17 \\
\bottomrule
\end{tabular}
\end{table}

In addition to assessing goodness of fit, we also evaluate the variable selection performance of each method across the different scenarios. Table \ref{tab:sim2_z_sofr} reports the proportion of the 100 simulated datasets in which each covariate is selected for each method under the different scenarios. With exception of grLASSO, all methods demonstrate strong performance in identifying the correct set of relevant covariates, while excluding the non-relevant ones, with BGLSS showing perfect selection performance across all scenarios, and our VEM algorithm likewise achieves perfect selection, with the exception of one dataset generated under the larger sample size scenario. In general, selection improves with larger sample sizes and smaller error variances, as indicated by the decrease in the proportion of datasets in which the non-relevant covariates (Covariates 2 and 4) are incorrectly included in the model.

\begin{figure}[ht]
    \centering
    \begin{subfigure}[b]{0.48\textwidth}
        \centering
        \includegraphics[width=\textwidth, page = 1]{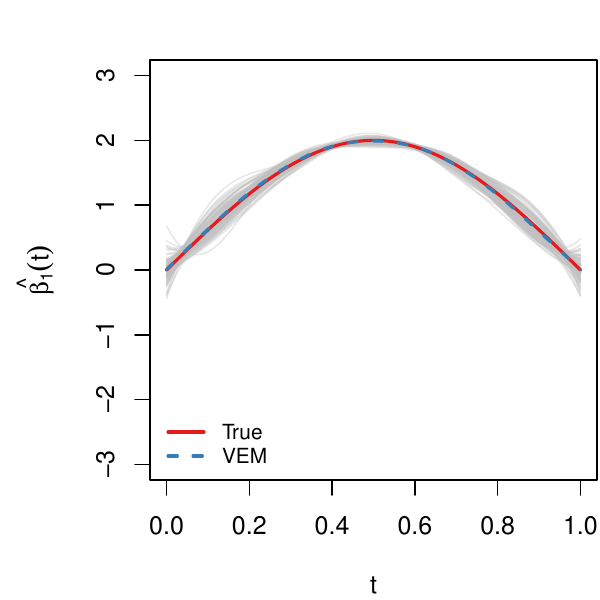}
        \includegraphics[width=\textwidth, page = 3]{sim2_vem_n400_sigma20.01.pdf}
        \caption{$\sigma^2 = 0.01$}
    \end{subfigure}
    \hfill
    \begin{subfigure}[b]{0.48\textwidth}
        \centering
        \includegraphics[width=\textwidth, page = 1]{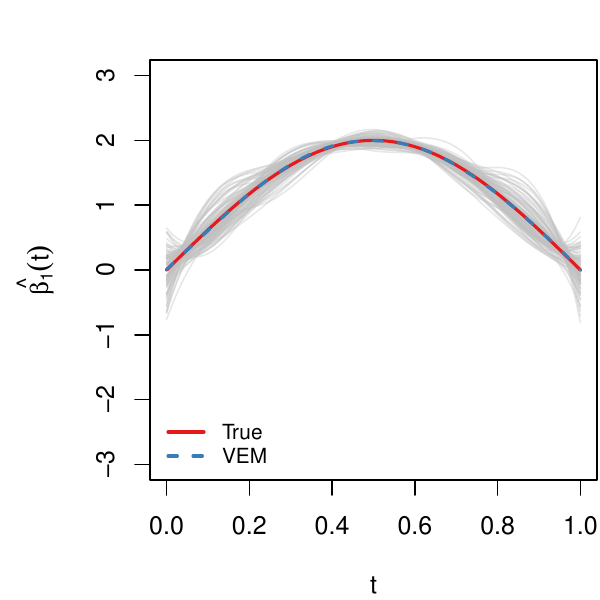}
        \includegraphics[width=\textwidth, page = 3]{sim2_vem_n400_sigma20.05.pdf}
        \caption{$\sigma^2 = 0.05$}
    \end{subfigure}
    \caption[Simulation Study 2. Estimated curves for the non-zero functional coefficients per dataset for VEM when $n = 400$.]{Simulation Study 2. Mean estimated curves for the non-zero partial functional coefficients obtained from our method (blue) for sample size $n = 400$ and varying error variance ($\sigma^2 = 0.01$ for plots on the left, and  $\sigma^2 = 0.05$ for plots on the right). The true values of the partial functional coefficients are shown in red and individual estimates curves across the 100 simulated datasets obtained from our method are provided in grey.}
    \label{fig:sim2_curve_sofr}
\end{figure}

Figure \ref{fig:sim2_curve_sofr} presents the mean estimated curves for the non-zero partial functional coefficients obtained from our method for scenarios with sample size $n = 400$, along with the individual estimated curves from our VEM algorithm across all simulations. Results from our method under the other scenarios and from the other methods are provided in Section 2 of the Supplementary Material. Note that, with exception of BGLSS, the estimation performance of the other methods were poor, as indicated by EMISE values in Table \ref{tab:sim2est_sofr}. The mean estimated curves are computed as described beforehand, we note that the mean estimated curves for the partial functional coefficients associated to the relevant covariates obtained from our method closely align with the true functional coefficients. We also note that the variance in the estimated functions obtained from our method decreases as the error variance decreases. In addition, all methods accurate estimate the intercept, regardless of the error variance (results not shown).

\begin{figure}[ht]
\centering
\begin{subfigure}[b]{0.48\textwidth}
    \centering
    \includegraphics[width=\textwidth, page = 1]{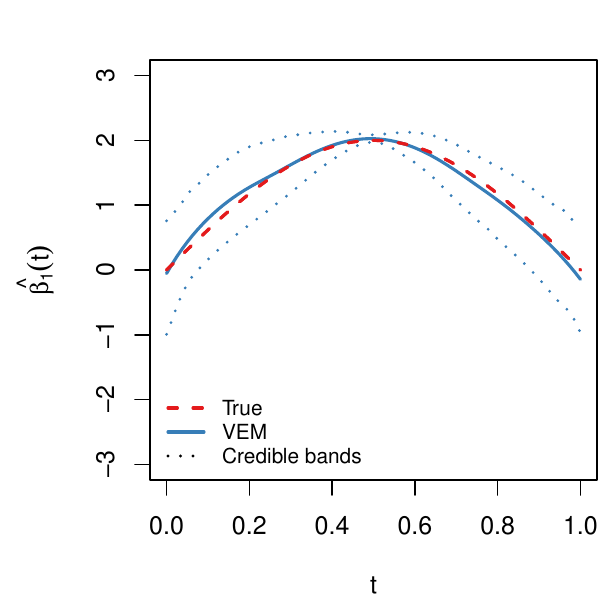}
    \includegraphics[width=\textwidth, page = 3]{sim2_vem_data97_n400_sigma20.01_cred.pdf}
    \caption{$\sigma^2 = 0.01$}
\end{subfigure}
\hfill
\begin{subfigure}[b]{0.48\textwidth}
    \centering
    \includegraphics[width=\textwidth, page = 1]{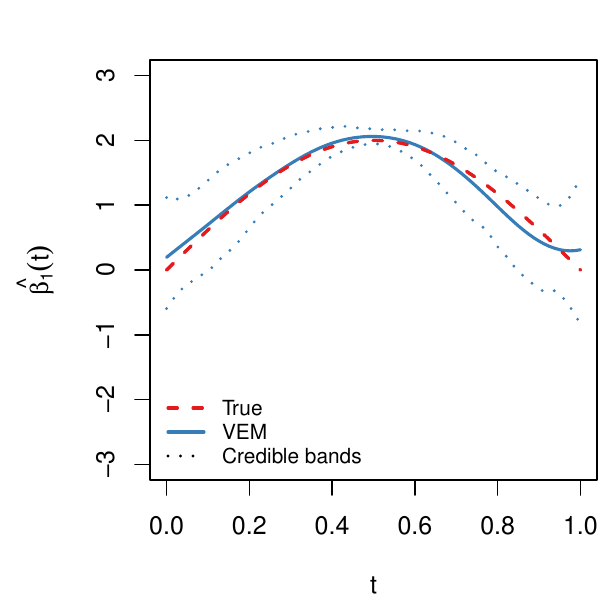}
    \includegraphics[width=\textwidth, page = 3]{sim2_vem_data97_n400_sigma20.05_cred.pdf}
    \caption{$\sigma^2 = 0.05$}
\end{subfigure}

\caption[Simulation Study 2. Estimated curves for non-zero partial functional coefficients with credible bands for VEM when $n = 400$.]{Simulation Study 2. Estimated curves for the non-zero partial functional coefficients from our VEM algorithm (blue) for one of the simulated datasets under the scenario with sample size $n = 400$ and error variances ($\sigma^2 = 0.01$ for plots on the left, and  $\sigma^2 = 0.05$ for plots on the right). A 95\% credible band (dotted curves) was obtained for our method. The true partial functional coefficients are shown in red.}
\label{fig:sim2_cred_sofr}
\end{figure}

In Figure \ref{fig:sim2_cred_sofr}, we present the estimated non-zero functional coefficients obtained from our method for one of the simulated datasets for scenarios with $n = 400$. A 95\% credible band was constructed as described in Simulation Study 1. The credible band cover the true functional coefficient curves across the entire domain and, as expected, become narrower with decreasing error variance. 


Overall, we note that grLASSO, grMCP and grSCAD are unable to effectively balance accuracy in the selection of the functional covariates, while preserving quality of fit. That is, these methods would need to lose fit quality and increase regularization to improve the accuracy in the selection, specially grLASSO. Our VEM algorithm and BGLSS are competitive in quality of fit and variable selection performance. Notably, our algorithm is on average 20 and 18 times faster than BGLSS in the scenarios with $n = 100$ and $n = 400$, respectively.

\subsubsection{Simulation Study 3}

In Simulation Study 3, we evaluate the performance of our method in terms of goodness of fit and variable selection in a partially functional regression model. Table \ref{tab:sim3est_sofr} summarizes the predictive and estimation performances across all scenarios and simulated datasets, using the MSE and EMISE metrics. The MSE values are averaged over the 100 simulated datasets. Overall, our VEM algorithm demonstrates strong goodness of fit, as reflected by the MSE values, which improve as the sample size and error variance decrease.

In terms of estimation performance, our VEM algorithm accurately estimates both functional and scalar coefficients. Table \ref{tab:sim3est_sofr} indicates that the estimation of final functional coefficients improves with larger sample sizes and smaller error variances, as shown by the decreasing EMISE values for the relevant covariate and for the zero values obtained for the non-relevant covariate. To evaluate the estimation of the scalar coefficients, we compute the average MSE for each scalar coefficient across the 100 datasets. Specifically, the estimate of the $l$th scalar coefficient is computed as $\hat\alpha_l = \frac{1}{100}\sum_{s = 1}^{100}\mu_{\alpha_l}^s$, where $\mu_{\alpha_l}^s$ denotes the estimated mean of the variational distribution of the $l$th scalar coefficient in the $s$th simulated dataset. Our method exhibits strong performance in estimating the scalar coefficients, with average MSE values of zero under the smaller error variance scenario and a maximum average MSE of only 0.003 for $\alpha_2$, corresponding to the relevant covariate, under the larger error variance scenario and smaller sample size. It is important to note that, as in Simulation Study 1, varying the error variance changes only the scale of the non-zero functional coefficient, while its shape remains the same, allowing results to be compared across scenarios.

\begin{table}[ht]
\centering
\caption[Simulation Study 3. MSE and EMISE values.]{Simulation Study 3. Mean squared error (MSE) values to assess goodness of fit taken as the average across the 100 simulated datasets for each scenario evaluated at different error variances ($\sigma^2 = 0.1, 0.5$) and sample sizes ($n = 50, 100$). The empirical mean integrated squared error (EMISE) is used to evaluate the estimation of the final functional coefficients and is computed as in Equation \eqref{eq:EMISE}.}
\label{tab:sim3est_sofr}
\begin{tabular}{lcc cc}
\toprule
 & \multicolumn{2}{c}{$\sigma^2 = 0.1$} & \multicolumn{2}{c}{$\sigma^2 = 0.5$} \\
\cmidrule(lr){2-3} \cmidrule(lr){4-5}
 & $n=50$ & $n=100$ & $n=50$ & $n=100$ \\
\midrule
MSE & 9.1992 & 7.9151 & 31.0574 & 23.3210 \\ 
\midrule
EMISE$_1$ & 5.9220 & 3.3145 & 19.6980 & 13.8079 \\ 
EMISE$_2$ & 0.0000 & 0.0000 & 0.0000 & 0.0000 \\ 
\bottomrule
\end{tabular}
\end{table}

Table \ref{tab:sim3z_sofr} presents the proportion of datasets in which each covariate (functional and scalar) is included in the model across the 100 simulated datasets. Functional Covariate 1 and scalar Covariate 2, which are the truly relevant covariates, are consistently selected across all scenarios. The non-relevant scalar covariate is incorrectly selected in at most 9\% of the simulated datasets under the larger-variance scenario, while the non-relevant functional covariate $X_1(t)$
is correctly excluded in all scenarios.

\begin{table}[ht]
\centering
\caption[Simulation Study 3. Covariate selection performance.]{Simulation Study 3. Proportion of datasets each covariate is selected across the 100 simulated datasets for each scenario evaluated at different error variances ($\sigma^2 = 0.1, 0.5$) and sample sizes ($n = 50, 100$).}
\label{tab:sim3z_sofr}
\begin{tabular}{lrrrr}
\toprule
&  \multicolumn{2}{c}{$\sigma^2 = 0.1$} & \multicolumn{2}{c}{$\sigma^2 = 0.5$} \\
\cmidrule(lr){2-3} \cmidrule(lr){4-5}
  &  $n = 50$ & $n = 100$ &  $n = 50$ & $n = 100$ \\
\midrule
Functional Covariate 1 & 1.00 & 1.00 & 1.00 & 1.00 \\ 
Functional Covariate 2 & 0.00 & 0.00 & 0.00 & 0.00 \\ 
\midrule
Scalar Covariate 1 & 0.06 & 0.07 & 0.09 & 0.09 \\ 
Scalar Covariate 2 & 1.00 & 1.00 & 1.00 & 1.00 \\ 
\bottomrule
\end{tabular}
\end{table}

\begin{figure}[ht]
    \centering
     \begin{subfigure}[b]{0.38\textwidth}
        \centering
        \includegraphics[width=\textwidth, page = 1]{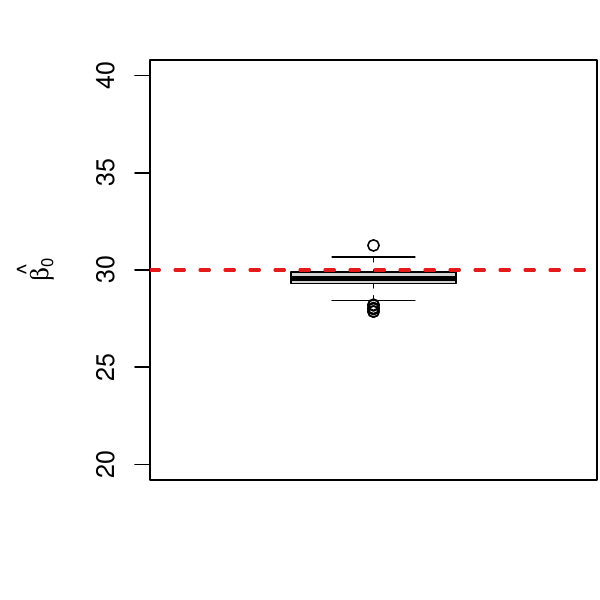}
        \includegraphics[width=\textwidth, page = 2]{sim3_vem_n100_sigma20.1.pdf}
        \caption{$\sigma^2 = 0.1$}
    \end{subfigure}
    \begin{subfigure}[b]{0.38\textwidth}
        \centering
        \includegraphics[width=\textwidth, page = 1]{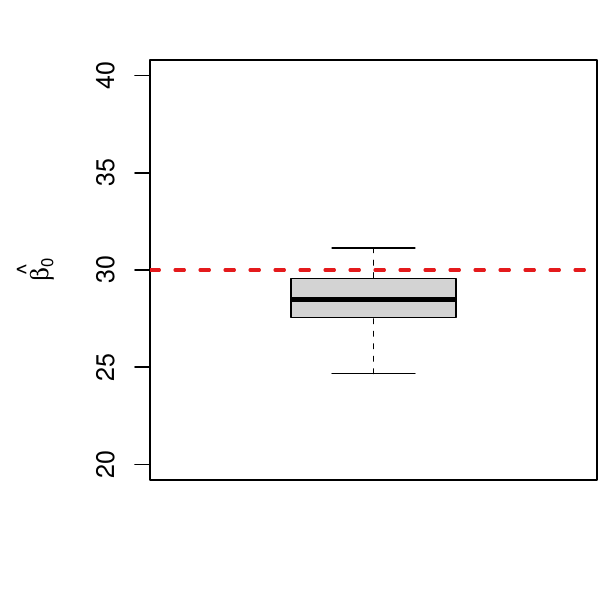}
        \includegraphics[width=\textwidth, page = 2]{sim3_vem_n100_sigma20.5.pdf}
        \caption{$\sigma^2 = 0.5$}
    \end{subfigure}
    \caption[Simulation Study 3. Estimated non-zero partial functional coefficient per dataset and boxplot of the intercept when $n = 100$.]{Simulation Study 3. Mean estimated curves (blue) for the non-zero partial functional coefficient, individual estimated curves for each simulated dataset (grey) and boxplot of the estimated intercept across 100 simulated datasets, for sample size $n = 100$ and varying error variance ($\sigma^2 = 0.1$ for plots on the left, and  $\sigma^2 = 0.5$ for plots on the right). The true values of the partial functional coefficient and intercept are shown in red.}
    \label{fig:sim3_curves_sofr}
\end{figure}

Figure \ref{fig:sim3_curves_sofr} presents the mean estimated curves for the non-zero partial functional coefficient and individual estimated curves for each simulated dataset for scenarios with $n = 100$. Results for the other scenarios are presented in Section 2 of the Supplementary Material. The mean estimated curves for the non-zero partial functional coefficient satisfactorily align with the true curves at most evaluation points, with minor deviations at the valleys. The functional coefficient associated to the non-relevant covariate is estimated as zero across all scenarios. Regarding the estimation of the intercept, we note that a slightly underestimation probably due to the fact that the functional coefficient associate to the relevant covariate is also being underestimated in some regions.

Complementing this, Figure 11 in Section 2 of the Supplementary Material shows the estimated non-zero partial functional coefficient with their corresponding  95\% credible bands for one of the simulated datasets for scenarios with $n = 100$. It is important to note that, for the simulated dataset considered, the true curves of the non-zero functional coefficient do not lie entirely within the credible bands across the whole evaluation domain. However, in some regions, they are contained within the bands, showing a better coverage in the scenario with smaller error variance. 


\clearpage
\section{Real data analysis}
\label{sec:real_data_sofr}

In this section, we apply our proposed method and compare its performance with the alternative approaches grLASSO, grMCP, grSCAD and BGLSS in the analyses of the sugar spectra and the Japan weather datasets. In both analyses, we standardize the functional covariates and center the response variable prior to applying our method and the alternative approaches. For covariate selection, as in the simulation studies, a covariate is selected by our method when the mode of the variational distribution of its corresponding $Z_j$ is one, which is equivalent to using a threshold of 0.5 on its inclusion probability $pz_j$, whereas for the other methods, a covariate is selected if at least one of its estimated basis coefficients is non-zero. To assess predictive performance across methods, we use the following metric, which is similar to the adjusted $R^2$, defined as
\begin{equation}
R^2_a =  1 - \frac{(n - 1)\sum_{i = 1}^{n}(y_i - \hat{y}_i)^2}{\left[n - \left(\sum_{j = 1}^{p}I_{\{\hat{Z}_j > 0\}}\right) \times K\right] \Big(\sum_{i=1}^n(y_i - \bar{y})^2\Big)},
\end{equation}
where $\hat{Z}_j$ represents the estimated latent variable indicating whether the $j$th covariate is included in the model. This metric accounts for the number of selected covariates, with values closer to one indicating better performance.

\subsection{Sugar spectra dataset}
\label{sec:sugar}

\begin{figure}[ht]
    \centering
    \includegraphics[width=\textwidth]{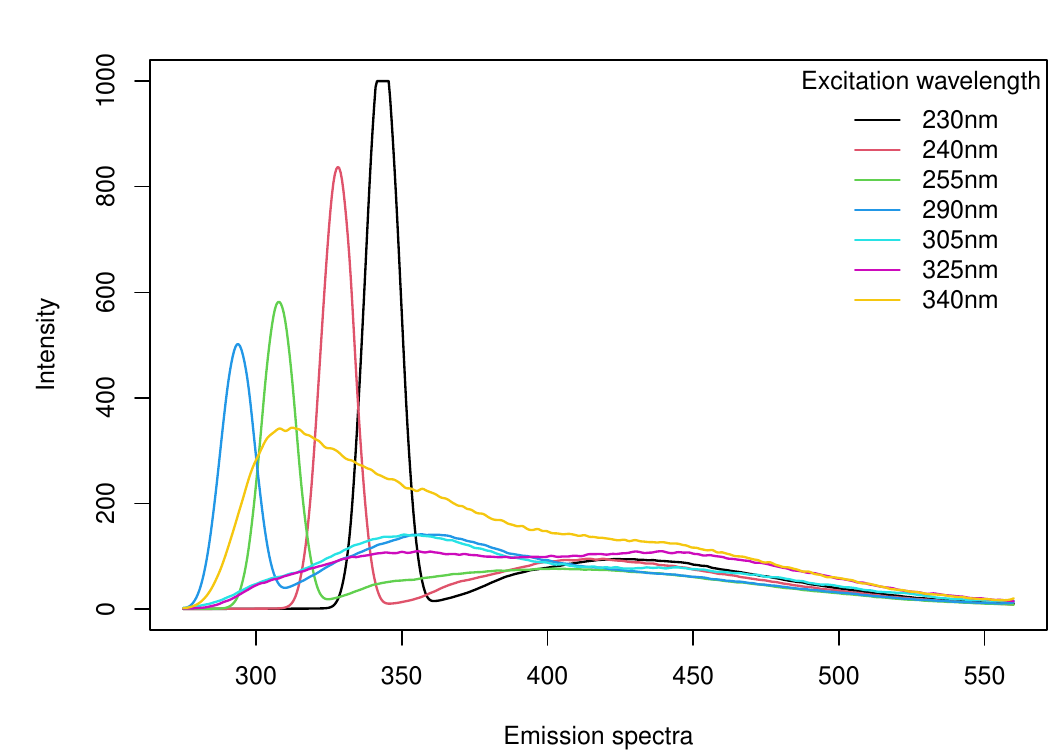}
    \caption[Sugar spectra dataset. Fluorescence spectra from a sugar sample.]{Sugar spectra dataset. Fluorescence spectra from one of the sugar samples at seven excitation wavelengths: 230, 240, 255, 290, 305, 325, 340 nm. The $y$-axis denotes the intensity and the $x$-axis represents the emission wavelengths from 275 nm to 560 nm measured in 0.5 nm intervals.}
    \label{fig:sugar_curves}
\end{figure}

The sugar spectra dataset, publicly available\footnote{Retrieved January 24, 2025, from \myurl{http://www.models.life.ku.dk/Sugar_Process}.}, was first introduced in \citet{munck1998} and \citet{bro1999} and later analysed by different works in the literature \citep{gertheiss2013,smaga2018,pannu2022}. It consists of 268 sugar samples analysed using chemometrics. For each sample, the emission spectra from 275 to 560 were measured in 0.5 nm intervals (571 measurements) at seven excitation wavelengths (230, 240, 255, 290, 305, 325, 340 nm). In addition to the fluorescence spectra, the ash content (in percentage), which measures the
amount of inorganic impurities in the refined sugar is also included. Raw absorbance curves for each excitation wavelength are provided in Figure \ref{fig:sugar_curves}.

In this analysis, we investigate the relationship between ash content and seven fluorescence spectra to identify the most relevant excitation wavelengths for predicting ash content. Fluorescence emission spectra measured at the seven excitation wavelengths are treated as functional covariates in the analysis.

To determine the optimal number and set of basis functions for representing the functional predictors and their corresponding coefficients, we employ our VEM algorithm using $K = 5, 6, 10, 12$ cubic B-splines basis functions. For each $K$, 50 random initializations of the inclusion probabilities $pz_j, \; j = 1, \dotsc, 7$, are considered, where each $pz_j$ is randomly assigned a value of either 0 or 1. The initialization leading to the VEM run with the highest ELBO is retained. Using the VEM run with the optimal initialization, we compute the generalized cross-validation (GCV) criterion \citep{golub1979}, and the optimal $K$ is selected according to the elbow plot procedure described by \citet{desouza2020}, adapting it to our situation. The elbow corresponds to the point on the GCV curve that lies farthest from the line connecting its first and last points.

Following the elbow plot procedure (not shown here), we found the set with $K = 6$ cubic B-splines to be the optimal set of basis functions. We apply all methods to the standardized sugar spectra dataset using the optimal set of $K = 6$ B-splines. The initialization of the parameter $\delta_2^*$ in the variational distribution of the error variance is set in such way that the average of the distribution is equal to the variance of the ash content, while the variance of the distribution is set to be reasonably large. The initialization of the other parameters in the VEM algorithm are set as described in Section \ref{sec:sofr_ini}. As in the simulation studies, the results for grLASSO, grMCP and grSCAD are obtained using the \R package \pck{grpreg}, with the regularization parameter being selected via cross validation. For the BGLSS, we run 10,000 iterations with a burn-in of 5000 samples using the implemented Gibbs sampler in the \R package \pck{MBSGS}.

Table \ref{tab:sugar_z} presents the set of excitation wavelengths selected by each method. We note that all methods selected the excitation wavelengths 325 nm and 340 nm, with exception of grSCAD. In addition, grLASSO, grMCP, and BGLSS provide the least parsimonious solutions. Specifically, grLASSO fails to eliminate any excitation wavelengths from the initial set, grMCP removes only the 290 nm wavelength, and BGLSS excludes only the 305 nm wavelength. In contrast, our VEM algorithm and grSCAD produce the sparsest solutions, each retaining only three covariates from the original set of seven fluorescence spectra.

\begin{table}[ht]
\centering
\caption[Sugar spectra dataset. Covariates selected for each method.]{Sugar spectra dataset. Excitation wavelengths of emission spectra selected as associated to ash content by our VEM algorithm, grLASSO, grMCP, grSCAD and BGLSS. Each entry in the table corresponds to a spectrum excitation wavelength (in nm) included in the model. All methods considered the same initial set of $K = 6$ cubic B-splines basis functions. }
\label{tab:sugar_z}
\begin{tabular}{ll}
\toprule
Method & Covariates included in the model\\
\midrule
VEM &  290, 325, 340\\
grLASSO & 230, 240, 255, 290, 305, 325, 340\\
grMCP & 230, 240, 255, 305, 325, 340 \\
grSCAD & 255, 290, 305\\
BGLSS & 230, 240, 255, 290, 325, 340\\
\bottomrule
\end{tabular}
\end{table}

\begin{figure}[ht]
    \centering
    \begin{subfigure}[b]{0.48\textwidth}
        \centering
        \includegraphics[width=\textwidth, page = 4]{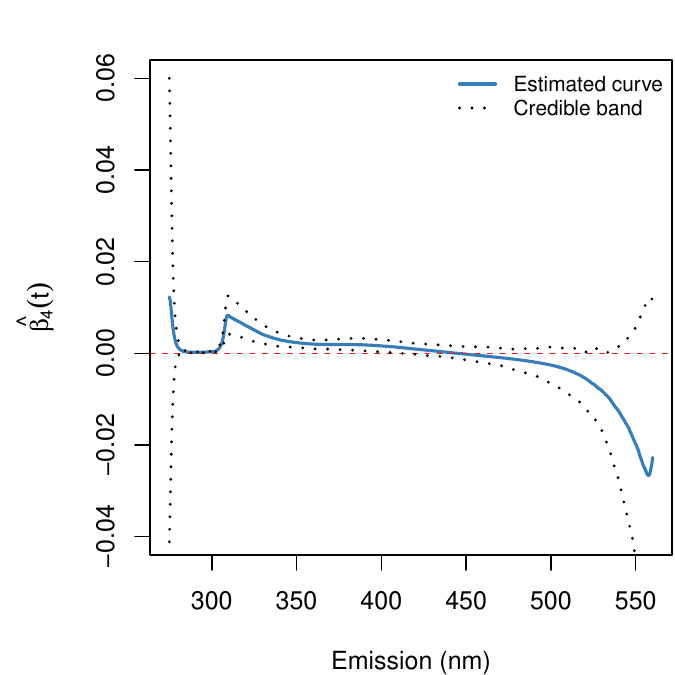}
        \caption{Excitation 290 nm}
    \end{subfigure}
    \begin{subfigure}[b]{0.48\textwidth}
        \centering
        \includegraphics[width=\textwidth, page = 6]{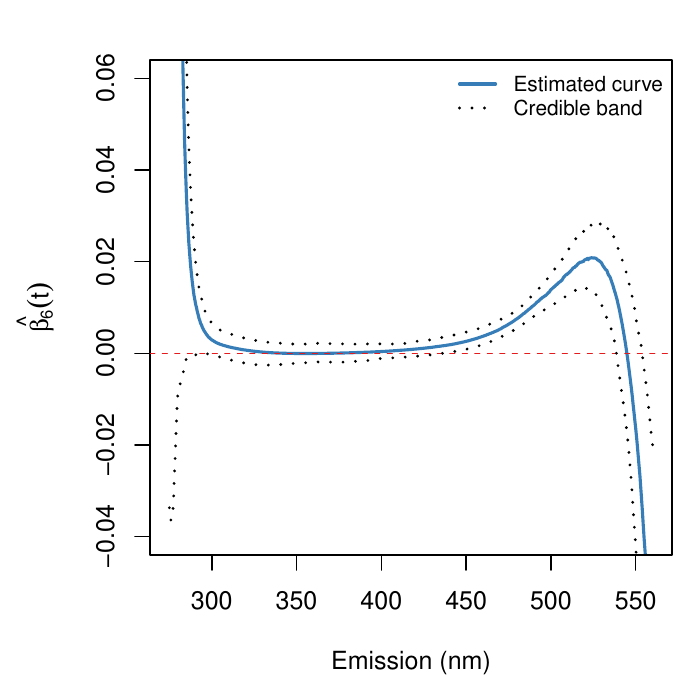}
        \caption{Excitation 325 nm}
    \end{subfigure}
    \begin{subfigure}[b]{0.48\textwidth}
        \centering
        \includegraphics[width=\textwidth, page = 7]{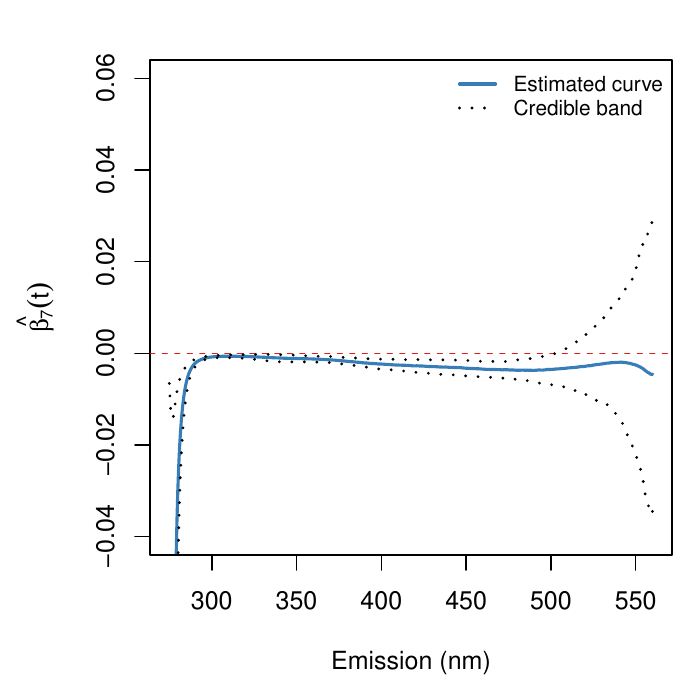}
        \caption{Excitation 340 nm}
    \end{subfigure}
    \caption[Sugar spectra dataset. Estimated curves for the non-zero functional coefficients for VEM.]{Sugar spectra dataset. Estimated curves (blue) for the functional coefficients associated to the selected excitation wavelengths, 290, 325, 340 nm, using our proposed method, with corresponding 95\% credible bands (dotted curves). A horizontal red line at zero is included for reference.}
    \label{fig:sugar_betas_vb}
\end{figure}

Figure \ref{fig:sugar_betas_vb} presents the estimated functional coefficients for the selected excitation wavelengths by our proposed method, along with their corresponding 95\% credible bands, computed as described in Section \ref{sec:sim_res_sofr}. The estimated partial functional coefficients associated with excitation wavelengths not included in the model were estimated as zero over the entire emission spectrum (results not shown). The estimated partial functional coefficients obtained from the other methods are included in Section 3 of the Supplementary Material.

Regarding predictive performance, we compare the methods based on the adjusted-$R^2$ metric, which takes into account the number of covariates selected by each method. We obtain the adjusted-$R^2$ values of 0.8464, 0.8572, 0.8540, 0.8409, 0.6586 for VEM, grLASSO, grMCP, grSCAD, and BGLSS, respectively. Notably, the methods show similar goodness-of-fit results, with grLASSO, grMCP, and our VEM algorithm achieving the highest predictive performance. However, grLASSO and grMCP provide the least parsimonious solutions. Therefore, our VEM algorithm provides the best balance between sparsity and predictive performance. In contrast, BGLSS shows the least predictive performance.

\subsection{Japan weather dataset}

Our second application involves publicly available\footnote{Retrieved August 23, 2025, from \myurl{http://www.data.jma.go.jp/obd/stats/data/en/}.} weather data collected from 83 weather stations in Japan. The dataset contains monthly and total weather measurements averaged over the period of 1991 to 2020 and has been analysed in the context of variable selection in functional regression by \citet{matsui2011}, \citet{collazos2016}, \citet{pannu2022} and \citet{mbina2025}. It should be noted, however, that the dataset used in \citet{matsui2011}, \citet{collazos2016} and \citet{pannu2022} differs from ours, as their analysis was based on data from 79 stations covering the period 1971 to 2000. Our dataset is more closely aligned with that of \citet{mbina2025}, as they also consider the period of 1991 to 2020, though we extend the analysis from 79 to 83 stations.

We consider average temperature (TEMP), average daylight duration (LIGHT), average relative humidity (HUMID), average maximum temperature (MAX.TEMP), average minimum temperature (MIN.TEMP) and average atmospheric pressure (PRESS) as functional predictors, which are measured monthly. The response variable is the average annual total precipitation (ATP) and we are interested in determining which monthly weather variables are relevant for predicting ATP.

The initialization of our VEM algorithm follows the same procedure as in the sugar spectra dataset analysis. We consider different sets of cubic B-spline basis functions with $K = 6, 8, 10, 12$. For each $K$, we perform 50 random initializations of the inclusion probabilities and select the optimal initialization based on the ELBO. The optimal number of basis functions $K$, is then determined using the elbow procedure. In addition, the initial value of the parameter $\delta_2^*$ in the variational distribution of the error variance is set according to the scale of the data.

We found $K = 8$ as the optimal number of basis functions based on the elbow procedure described in Section \ref{sec:sugar}. We fit each method using the same set of basis functions, using the \R package \pck{grpreg} to obtain the results for grLASSO, grMCP and grSCAD, with the regularization parameter being selected via cross validation. We consider the Gibbs sampler implemented in the \R package \pck{MBSGS} to fit the BGLSS method, with 10,000 iterations and a burn-in of 5000 samples. Table \ref{tab:jma_z} presents the set of monthly weather measurements selected for each method. Notably, average daylight duration is selected by all methods. Average temperature and average relative humidity are also selected by all approaches, with exception of grMCP. Among the competing procedures, grLASSO yields the least parsimonious solution, as it includes all predictors in the model.

\begin{table}[ht]
\centering
\caption[Japan weather dataset. Covariates selected for each method.]{Japan weather dataset. Covariates selected as associated to average annual total precipitation by our VEM algorithm, grLASSO, grMCP, grSCAD, and BGLSS. All methods considered the same initial set of $K = 8$ cubic B-splines basis functions.}
\label{tab:jma_z}
\begin{tabular}{ll}
\toprule
Method & Covariates included in the model\\
\midrule
VEM &  TEMP, HUMID, LIGHT\\
grLASSO & TEMP, MAX.TEMP, MIN.TEMP, PRESSURE, HUMID, LIGHT\\
grMCP & PRESSURE, LIGHT\\
grSCAD & TEMP, PRESSURE, HUMID, LIGHT\\
BGLSS & TEMP, MAX.TEMP, HUMID, LIGHT\\
\bottomrule
\end{tabular}
\end{table}

\begin{figure}[ht]
    \centering
    \begin{subfigure}[b]{0.48\textwidth}
        \centering
        \includegraphics[width=\textwidth, page = 1]{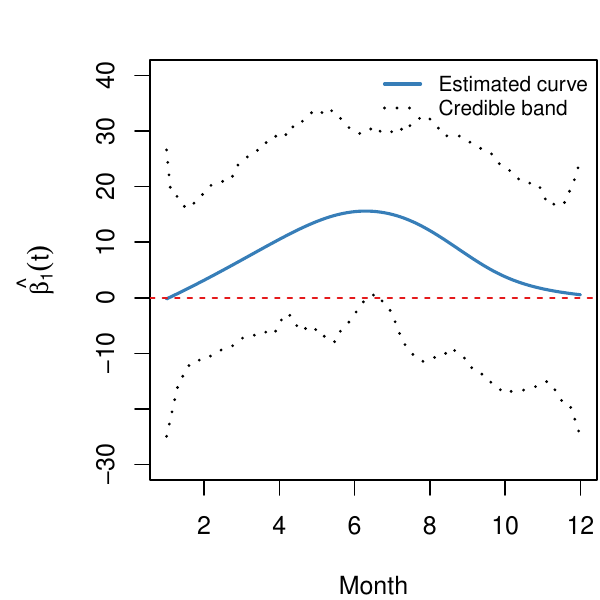}
        \caption{Temperature}
    \end{subfigure}
    \hfill
    \begin{subfigure}[b]{0.48\textwidth}
        \centering
        \includegraphics[width=\textwidth, page = 5]{JMA_res_VEM.pdf}
        \caption{Humidity}
    \end{subfigure}
    \begin{subfigure}[b]{0.48\textwidth}
        \centering
        \includegraphics[width=\textwidth, page = 6]{JMA_res_VEM.pdf}
        \caption{Daylight duration}
    \end{subfigure}
    \caption[Japan weather dataset. Estimated curves for the non-zero functional coefficients for VEM.]{Japan weather dataset. Estimated curves (blue) for the functional coefficients associated to the selected weather information by our proposed method: average temperature, average relative humidity and average total daylight duration, with corresponding 95\% credible band (dotted curves). A horizontal red line at zero is included for reference.}
    \label{fig:jma_betas_vb}
\end{figure}

We present in Figure \ref{fig:jma_betas_vb} the estimated partial functional coefficients for each predictor selected by our method with their corresponding 95\% credible bands, computed as described in Section \ref{sec:sim_res_sofr}. The estimated effect of the average monthly temperature on the average annual total precipitation is positive throughout the year, with a stronger association during the summer months. Similarly, humidity exhibits a consistently positive contribution, with greater effect towards the end of the summer and beginning of fall. In contrast, the effect of average daylight duration varies in its direction of association, contributing negatively to precipitation during certain months of the year. The estimated coefficients obtained from the other methods are included in Section 3 of the Supplementary Material.

In terms of predictive performance, the adjusted-$R^2$ values are 0.6011, 0.6093, 0.7393, 0.6820, and 0.3864, for VEM, grLASSO, grMCP, grSCAD, and BGLSS, respectively. Overall, the results are relatively comparable across methods, with grMCP showing the highest predictive performance, but still only explaining 74\% of the variability in the ash content. It is important to note, however, that the current analysis is based on a small number of observations, in contrast to the 268 samples used in the sugar spectra data analysis.

\clearpage

\section{Discussion}
\label{sec:conclusion_sofr}

In this work, we focus on estimation and variable selection in SoFR by developing VEM algorithms for scalar-on-function and partially functional regression models, to the best of our knowledge, our work stands as one of the first to employ a Bayesian variational inference approach to perform estimation and variable selection in these models. Our approach for variable selection using latent random variables under a Beta--Bernoulli prior specification can be employed to the other types of functional regression.

Through simulation studies, our method was capable of identifying the set of relevant covariates most of the time while maintaining an optimal level of goodness of fit. Compared with alternative linear regression approaches, our method achieves a better balance between goodness of fit and variable selection than grLASSO, grMCP, and grSCAD, and performs comparably to BGLSS. Both VEM algorithm and BGLSS demonstrate competitive performances, achieving high predictive accuracy and accurate covariate selection, which is expected given that both employ sparsity-inducing priors enabling adaptive selection. 

In the real data applications, our method performed variable selection while maintaining satisfactory goodness of fit in both applications, being on average eight times faster than BGLSS. Notably, alternative procedures could be considered to evaluate predictive ability, rather than relying only on adjusted-$R^2$ for comparison. If prediction is of interest, partitioning the data into training and test sets would provide a more informative assessment of the predictive performance of each method. For Bayesian approaches, evaluating the posterior predictive distribution \citep{neal2011, dias2013} could further inform predictive accuracy.

As future work, one could explore alternative procedures for computing credible bands, such as using a depth function \citep{cuevas2007, febrero2012}, and develop a metric to evaluate the proposed band coverage across different datasets. In addition, our framework may be extended to repeated measurements across different groups by including group-specific random intercepts to account for the within-group correlation. 
Our method can be further extended to non-Gaussian responses, such as Bernoulli and Poisson, enabling adaptive selection of both scalar and functional covariates across a broader class of functional regression models. Extensions to time-to-event outcomes, such as survival models, also represent a promising direction for future work.


\clearpage

\appendix
\input{supp_sofr_vbelbo}

\bibliographystyle{apalike}
\bibliography{ref_sofr}
\end{document}

%% file: supp_sofr_vbelbo.tex
\section{Expectations and ELBO derivation}
\label{ap:vb_exp}

In Appendix \ref{ap:vb_exp}, we present the expectations used in the VB updated equations derived in Section \ref{sec:vb_sofr} and derive the ELBO, the convergence criterion used in our VEM algorithm.

\subsection{Expectations}
\label{sec:vb_exp_sofr}

\begin{gather*}
\E_{q(\sigma^2)}\left(\frac{1}{\sigma^2}\right) = \frac{\delta_1^*}{\delta_2^*};
\\
\E_{q(\sigma^2)}(\log{\sigma^2}) = \log\delta_2^* - \mathrm{digamma}({\delta_1^*});
\\
\E_{q(\theta_j)}(\log{\theta_j}) = \mathrm{digamma}(a_j) - \mathrm{digamma}(a_j + b_j);
\\
\E_{q(\theta_j)}(\log(1-{\theta_j})) = \mathrm{digamma}(b_j) - \mathrm{digamma}(a_j + b_j);
\\
\E_{q(\tau^2_{kj})}(\tau^2_{kj}) =
\left(\frac{\chi_{kj}}{\psi_{kj}}\right)^{1/2}
\frac{\mathcal{K}_{3/2}(\sqrt{\psi_{kj}\chi_{kj}})}
{\mathcal{K}_{1/2}(\sqrt{\psi_{kj}\chi_{kj}})};
\\
\E_{q(\tau^2_{kj})}\left(\frac{1}{\tau^2_{kj}}\right) =
\left(\frac{\psi_{kj}}{\chi_{kj}}\right)^{1/2}
\frac{\mathcal{K}_{3/2}(\sqrt{\psi_{kj}\chi_{kj}})}
{\mathcal{K}_{1/2}(\sqrt{\psi_{kj}\chi_{kj}})} -\frac{1}{\chi_{kj}};
\text{and}
\\
\E_{q(\mathbf{Z})}(\Gamma) = \diag(p_{\mathbf{Z}}),
\end{gather*}
where $\mathcal{K}$ is the modified Bessel function of the second kind and $p_\mathbf{Z} = (pz_1^*, \dotsc, pz_p^*)$. The expectations with respect to $\tau^2_{kj}$ are computed using the \fn{Egig} function from the \R package \pck{lqr} \citep{lqr2024}.

Since $\Gamma$ is a diagonal matrix, $\Gamma W\trp W \Gamma = (W\trp W)\odot(\mathbf{Z}\mathbf{Z}\trp)$. We can also show that $\E_{q(\mathbf{Z})}(Z_i Z_j) = pz_i^* pz_j^* \;(i \neq j)$ and $\E_{q(\mathbf{Z})}(Z_j^2) = pz_j^* = pz_j^{{*}^2} + pz_j^*(1-pz_j^*)$. We can then obtain that
\begin{equation*}
    \E_{q(\mathbf{Z})}(\Gamma W\trp W \Gamma) = (W\trp W) \odot \Omega,
\end{equation*}
where $\Omega = p_\mathbf{Z}p_\mathbf{Z}\trp + \diag(p_{\mathbf{Z}})(\mathrm{I} - \diag(p_{\mathbf{Z}}))$, and $\odot$ represents Hadamard product between two matrices.

\begin{align*}
\E_{q(\mathbf{Z})q(\mathbf{b})}(\mathbf{y} - W\Gamma\mathbf{b})\trp(\mathbf{y} - W\Gamma\mathbf{b}) =&\ \mathbf{y}\trp\mathbf{y} -2\mathbf{y}\trp W\diag(p_{\mathbf{Z}})\mu_{\mathbf{b}}
\\
& {} + \E_{q(\mathbf{b})q(\mathbf{Z})}(\mathbf{b}\trp\Gamma W\trp W \Gamma\mathbf{b}),
\nonumber
\end{align*}
where $\E_{q(\mathbf{b})q(\mathbf{Z})}(\mathbf{b}\trp\Gamma W\trp W \Gamma\mathbf{b}) = \mathrm{tr}\left\{(\Sigma_{\mathbf{b}} + \mu_{\mathbf{b}}\mu_{\mathbf{b}}\trp)(W\trp W \odot \Omega)\right\}$.

Finally, $\E_{q(\tau^2_{kj})}(\log \tau^2_{kj})$ is also computed using the \fn{Egig} function from the \R package \pck{lqr}.

\subsection{Evidence Lower Bound (ELBO)}
\label{sec:elbo_sofr}

We derive the ELBO for the scalar-on-function regression model with the corresponding Bayesian hierarchical model in Equation \eqref{eq:hier_sofr}.
\begin{equation*}
\mathrm{ELBO}(q) = \E_{q}(\log p(\mathbf{Y}, \mathbf{Z}, \mathbf{b}, \boldsymbol{\theta}, \sigma^2, \boldsymbol{\tau}^2, \boldsymbol{\lambda}^2)) - \E_{q}(\log q( \mathbf{Z}, \mathbf{b}, \boldsymbol{\theta}, \sigma^2, \boldsymbol{\tau}^2)).
\end{equation*}

Using the decomposition of the complete-data likelihood given in \eqref{eq:vb_complete_sofr} and the mean-field variational family assumption, we calculate the ELBO as follows:
\begin{align*}
\label{eq:elbo_sofr}
\mathrm{ELBO}(q)
=&\ \E_{q}(\log p(\mathbf{Y} \mid \mathbf{Z}, \mathbf{b}, \sigma^2))\\
& {} + \E_{q}(\log p(\mathbf{Z} \mid \boldsymbol{\theta})) -  \E_{q}(\log q( \mathbf{Z}))\\
& {} + \E_{q}(\log p(\mathbf{b} \mid \sigma^2, \boldsymbol{\tau}^2)) -  \E_{q}(\log q(\mathbf{b}) )\\
& {} + \E_{q}(\log p(\boldsymbol{\theta})) -  \E_{q}(\log q(\boldsymbol{\theta}))
\\
& {} + \E_{q}(\log p(\sigma^2)) -  \E_{q}(\log q(\sigma^2) )\\
& {} + \E_{q}(\log p(\boldsymbol{\tau}^2 \mid \boldsymbol{\lambda}^2)) -  \E_{q}(\log q(\boldsymbol{\tau}^2) ).
\end{align*}

Consequently, we define each term as follows:
\begin{align*}
&\ \E_{q}(\log p(\mathbf{Y} \mid \mathbf{Z}, \mathbf{b}, \sigma^2)) =
\\
=&\ \E_{q}\left\{-\frac{n}{2}\log(2\pi) -\frac{n}{2}\log(\sigma^2) -\frac{1}{2}\left(\frac{1}{\sigma^2}\right)(\boldsymbol{y} - W\Gamma\mathbf{b})\trp(\boldsymbol{y} - W\Gamma\mathbf{b})\right\}\\
=&\ - \frac{n}{2}\log(2\pi) -\frac{n}{2}\E_{q}(\log(\sigma^2)) -\frac{1}{2}\E_{q}\left(\frac{1}{\sigma^2}\right)\E_{q}[(\boldsymbol{y} - W\Gamma\mathbf{b})\trp(\boldsymbol{y} - W\Gamma\mathbf{b})].
\end{align*}

\begin{align*}
&\ \E_{q}(\log p(\mathbf{Z} \mid \boldsymbol{\theta})) - \E_{q}(\log q( \mathbf{Z})) =
\\
=&\  \E_{q}\left(\sum_{j=1}^p\log\left({\theta_j}^{Z_j}{(1-\theta_j)}^{(1-Z_j)}\right)\right) - \E_{q}\left(\sum_{j=1}^p\log\left({pz_j}^{Z_j}{(1-pz_j)}^{(1-Z_j)}\right)\right)\\
=&\ \E_{q}\left\{\sum_{j=1}^p\left(Z_j\left[\log\left(\frac{\theta_j}{1-\theta_j}\right) + \log\left(\frac{1-pz_j}{pz_j}\right)\right] + \log(1-\theta_j) - \log(1-pz_j)\right)\right\}\\
=&\ \sum_{j=1}^p \left(pz_j \left[\E_{q}\left(\log\left(\frac{\theta_j}{1-\theta_j}\right)\right) + \log\left(\frac{1-pz_j}{pz_j}\right)\right] + \E_{q}(\log(1-\theta_j)) - \log(1-pz_j)\right).
\end{align*}

\begin{align*}
&\ \E_{q}(\log p(\mathbf{b} \mid \sigma^2, \boldsymbol{\tau}^2)) -  \E_{q}(\log q(\mathbf{b})) =
\\
=&\ \E_{q}\left\{-\frac{Kp}{2}\log(2\pi) -\frac{Kp}{2}\log(\sigma^2)-\frac{1}{2}\sum_{j=1}^p\sum_{k=1}^K\log(\tau^2_{kj}) -\frac{1}{2}\left(\frac{1}{\sigma^2}\right)\sum_{j=1}^p\sum_{k=1}^K\left[\left(\frac{1}{\tau^2_{kj}}\right)b_{kj}^2\right]\right\}\\
& {} - \E_{q}\left\{-\frac{Kp}{2}\log(2\pi) -\frac{1}{2}\log(|{\Sigma_{\mathbf{b}}}|) - \frac{1}{2}(\mathbf{b} - \boldsymbol{\mu}_{\mathbf{b}})\trp \Sigma_{\mathbf{b}}^{-1}(\mathbf{b} - \boldsymbol{\mu}_{\mathbf{b}})\right\}\\
=&\ -\frac{Kp}{2}\E_{q}(\log(\sigma^2))-\frac{1}{2}\sum_{j=1}^p\sum_{k=1}^K \E_{q}(\log(\tau^2_{kj})) -\frac{1}{2}\E_{q}\left(\frac{1}{\sigma^2}\right)\sum_{j=1}^p\sum_{k=1}^K \left[\E_{q}\left(\frac{1}{\tau^2_{kj}}\right)\E_{q}(b_{kj}^2)\right]\\
& {} + \frac{1}{2}\log(|{\Sigma_{\mathbf{b}}}|)+ \frac{Kp}{2}.
\end{align*}
\begin{align*}
&\ \E_{q}(\log p(\boldsymbol{\theta})) -  \E_{q}(\log q(\boldsymbol{\theta})) = 
\\
=&\ \E_{q}\left\{\sum_{j=1}^p\log\left(\theta_j^{0.5 - 1}{(1-\theta_j)}^{0.5 - 1} \frac{\Gamma(1)}{\Gamma(0.5)\Gamma(0.5)}\right) \right\}\\
& {} - \E_{q}\left\{\sum_{j=1}^p\log\left(\theta_j^{a_j - 1}{(1-\theta_j)}^{b_j - 1} \frac{\Gamma(a_j + b_j)}{\Gamma(a_j)\Gamma(b_j)}\right)\right\}\\
=&\ \sum_{j=1}^p\left[(0.5 - 1)\E_{q}(\log(\theta_j)) + (0.5 - 1)\E_{q}(\log(1-\theta_j)) - \log(\Gamma(0.5)) - \log(\Gamma(0.5))\right]\\
& {} - \sum_{j=1}^p\left[(a_j - 1)\E_{q}(\log(\theta_j)) + (a_j - 1)\E_{q}(\log(1-\theta_j))  \right.\\
& {} + \left.\log(\Gamma(a_j + b_j)) - \log(\Gamma(a_j)) - \log(\Gamma(b_j))\right]\\
\approx&\ \sum_{j=1}^p\left[(0.5 - a_j)\E_{q}(\log(\theta_j)) + (0.5 - b_j)\E_{q}(\log(1-\theta_j)) \right.\\
& \left. {} - \log(\Gamma(a_j + b_j)) + \log(\Gamma(a_j)) + \log(\Gamma(b_j))\right].
\end{align*}
\begin{align*}
&\ \E_{q}(\log p(\sigma^2)) -  \E_{q}(\log q(\sigma^2)) =
\\
=&\ \E_{q}\left\{\delta_1\log\delta_2 - \log\Gamma(\delta_1) - (\delta_1 - 1)\log(\sigma^2) - \left(\frac{1}{\sigma^2}\right)\delta_2\right\}\\
& {} - \E_{q}\left\{\delta_1^*\log\delta_2^* - \log\Gamma(\delta_1^*) - (\delta_1^* - 1)\log(\sigma^2) - \left(\frac{1}{\sigma^2}\right)\delta_2^*\right\}\\
=&\ \delta_1\log\delta_2 - \log\Gamma(\delta_1) -\delta_1^*\log\delta_2^* + \log\Gamma(\delta_1^*) + (\delta_1^* - \delta_1)\E_{q}(\log(\sigma^2)) + (\delta_2^* - \delta_2)\E_{q}\left(\frac{1}{\sigma^2}\right). 
\end{align*}
\begin{align*}
&\ \E_{q}(\log p(\boldsymbol{\tau}^2 \mid \boldsymbol{\lambda}^2)) -  \E_{q}(\log q(\boldsymbol{\tau}^2)) =
\\
=&\ \E_{q}\left\{\sum_{j=1}^p\sum_{k=1}^K\left(\log(\lambda^2_j) -\log(2) -\frac{\lambda^2_j}{2}\tau^2_{kj}\right)\right\}\\
& {} - \E_{q}\left\{\sum_{j=1}^p\sum_{k=1}^K\left(\frac{1}{4}\log\left(\frac{\psi_{kj}}{\chi_{kj}}\right)^{1/4} - \log(2 \mathcal{K}_{1/2}(\sqrt{\chi_{kj}\psi_{kj}})) \right. \right.\\
& \left. \left. {} - \frac{1}{2}\log(\tau^2_{kj}) - \frac{1}{2}\left(\chi_{kj}\frac{1}{\tau^2_{kj}} + \psi_{kj}\tau^2_{kj}\right)\right)\right\}\\
=&\ K\sum_{j=1}^p \log(\lambda^2_j) -\frac{1}{2}\sum_{j=1}^p\sum_{k=1}^K(\lambda^2_j\E_{q}(\tau^2_{kj}))\\
& {} - \frac{1}{4}\sum_{j=1}^p\sum_{k=1}^K \left\{\log(\psi_{kj}) - \log(\chi_{kj}) - \log(\mathcal{K}_{1/2}(\sqrt{\chi_{kj}\psi_{kj}}))\right\}\\
& {} -\frac{1}{2}\sum_{j=1}^p\sum_{k=1}^K \E_{q}(\log(\tau^2_{kj})) -\frac{1}{2}\sum_{j=1}^p\sum_{k=1}^K\left(\chi_{kj} \E_{q}\left(\frac{1}{\tau^2_{kj}}\right) + \psi_{kj} \E_{q}(\tau^2_{kj})\right),
\end{align*}
where $\mathcal{K}_{1/2}$ is the modified Bessel function of the second kind.